\documentclass[prx,aps,twocolumn,superscriptaddress,showpacs,footinbib,longbibliography]{revtex4-1}
\usepackage[colorlinks=true,linkcolor=black, citecolor=blue, urlcolor=blue, 
    unicode=true]{hyperref}

\usepackage{tikz}
\usetikzlibrary{calc}
\usepackage{graphicx,amsmath,braket,tikz,amssymb,footmisc,lipsum,array,tabularx,float,comment,xcolor}

\usepackage{cleveref}
\newcommand{\mvo}{MgV$_{2}$O$_{4}$}

\crefname{equation}{}{}
\Crefname{equation}{}{}
\crefrangelabelformat{equation}{#3#1#4--#5#2#6}

\crefmultiformat{equation}{#2#1#3}{, #2#1#3}{#2#1#3}{#2#1#3}
\Crefmultiformat{equation}{#2#1#3}{, #2#1#3}{#2#1#3}{#2#1#3}
\begin{document}


\title{\texorpdfstring{Spin-orbital correlations from complex orbital order in MgV$_{2}$O$_{4}$}{}}

\author{H. Lane}
\affiliation{School of Physics and Astronomy, University of Edinburgh, Edinburgh EH9 3JZ, United Kingdom}
\affiliation{School of Chemistry and Centre for Science at Extreme Conditions, University of Edinburgh, Edinburgh EH9 3FJ, United Kingdom}
\affiliation{ISIS Pulsed Neutron and Muon Source, STFC Rutherford Appleton Laboratory, Harwell Campus, Didcot, Oxon, OX11 0QX, United Kingdom}
\affiliation{School of Physics, Georgia Institute of Technology, Atlanta, GA 30332, USA}

\author{P. M. Sarte}
\affiliation{Materials Department, University of California, Santa Barbara, CA 93106, USA}

\author{K. Guratinder}
\affiliation{School of Physics and Astronomy, University of Edinburgh, Edinburgh EH9 3JZ, United Kingdom}

\author{A. M. Arevalo-Lopez}
\affiliation{Univ. Lille, CNRS, Centrale Lille, Univ. Artois, UMR 8181 - UCCS - Unite de Catalyse et Chimie du Solide, F-59000 Lille, France}

\author{R. S. Perry}
\affiliation{ISIS Pulsed Neutron and Muon Source, STFC Rutherford Appleton Laboratory, Harwell Campus, Didcot, Oxon, OX11 0QX, United Kingdom}
\affiliation{London Centre for Nanotechnology and Department of Physics and Astronomy,
University College London, London WC1E 6BT, United Kingdom}

\author{E. C. Hunter}
\affiliation{School of Physics and Astronomy, University of Edinburgh, Edinburgh EH9 3JZ, United Kingdom}

\author{T. Weber}
\affiliation{Institute Laue-Langevin, 71 Avenue des Martyrs, CS 20156, 38042 Grenoble cedex 9, France}

\author{B. Roessli}
\affiliation{Laboratory for Neutron Scattering and Imaging, Paul Scherrer Institut (PSI), 5232 Villigen PSI, Switzerland}

\author{A. Stunault}
\affiliation{Institute Laue-Langevin, 71 Avenue des Martyrs, CS 20156, 38042 Grenoble cedex 9, France}

\author{Y. Su}
\affiliation{J\"{u}lich Centre for Neutron Science (JCNS) at Heinz Maier-Leibnitz Zentrum (MLZ), Forschungszentrum J\"{u}lich GmbH, Lichtenbergstra{\ss}e 1, 85748 Garching, Germany}

\author{R. A. Ewings}
\affiliation{ISIS Pulsed Neutron and Muon Source, STFC Rutherford Appleton Laboratory, Harwell Campus, Didcot, Oxon, OX11 0QX, United Kingdom}

\author{S.~D.~Wilson}
\affiliation{Materials Department and California Nanosystems Institute, University of California, Santa Barbara, CA 93106, USA}

\author{P. B\"oni}
\affiliation{Physik-Department E21, Technische Universit\"at M\"unchen (TUM), 
    James-Franck-Str. 1, 85748 Garching, Germany}

\author{J.~P.~Attfield}
\affiliation{School of Chemistry and Centre for Science at Extreme Conditions, University of Edinburgh, Edinburgh EH9 3FJ, United Kingdom}

\author{C. Stock}
\affiliation{School of Physics and Astronomy, University of Edinburgh, Edinburgh EH9 3JZ, United Kingdom}

\date{\today}

\begin{abstract}
MgV$_{2}$O$_{4}$ is a spinel based on magnetic V$^{3+}$ ions which host both spin ($S=1$) and orbital ($l_{eff}=1$) moments.  Owing to the underlying pyrochlore coordination of the magnetic sites, the spins in MgV$_{2}$O$_{4}$ only antiferromagnetically order once the frustrating interactions imposed by the $Fd\overline{3}m$ lattice are broken through an orbitally-driven structural distortion at T$_{S}$ $\simeq$ 60 K. Consequently, a N\'eel transition occurs at T$_{N}$ $\simeq$ 40 K.  Low temperature spatial ordering of the electronic orbitals is fundamental to both the structural and magnetic properties, however considerable discussion on whether it can be described by complex or real orbital ordering is ambiguous.  We apply neutron spectroscopy to resolve the nature of the orbital ground state and characterize hysteretic spin-orbital correlations using x-ray and neutron diffraction.  Neutron spectroscopy finds multiple excitation bands and we parameterize these in terms of a multi-level (or excitonic) theory based on the orbitally degenerate ground state.  Meaningful for the orbital ground state, we report an ``optical-like" mode at high energies that we attribute to a crystal-field-like excitation from the spin-orbital $j_{eff}$=2 ground state manifold to an excited $j_{eff}$=1 energy level.  We parameterize the magnetic excitations in terms of a Hamiltonian with   spin-orbit coupling and local crystalline electric field distortions resulting from deviations from perfect octahedra surrounding the V$^{3+}$ ions.  We suggest that this provides compelling evidence for complex orbital order in MgV$_{2}$O$_{4}$. We then apply the consequences of this model to understand hysteretic effects in the magnetic diffuse scattering where we propose that MgV$_{2}$O$_{4}$ displays a high temperature orbital memory of the low temperature spin order.

\end{abstract}

\pacs{}

\maketitle

\section{Introduction}
\label{Sect:Introduction}

\begin{figure}
	\begin{center}
		\includegraphics[width=\linewidth,trim=0mm 20mm 0mm 0mm, clip=true]{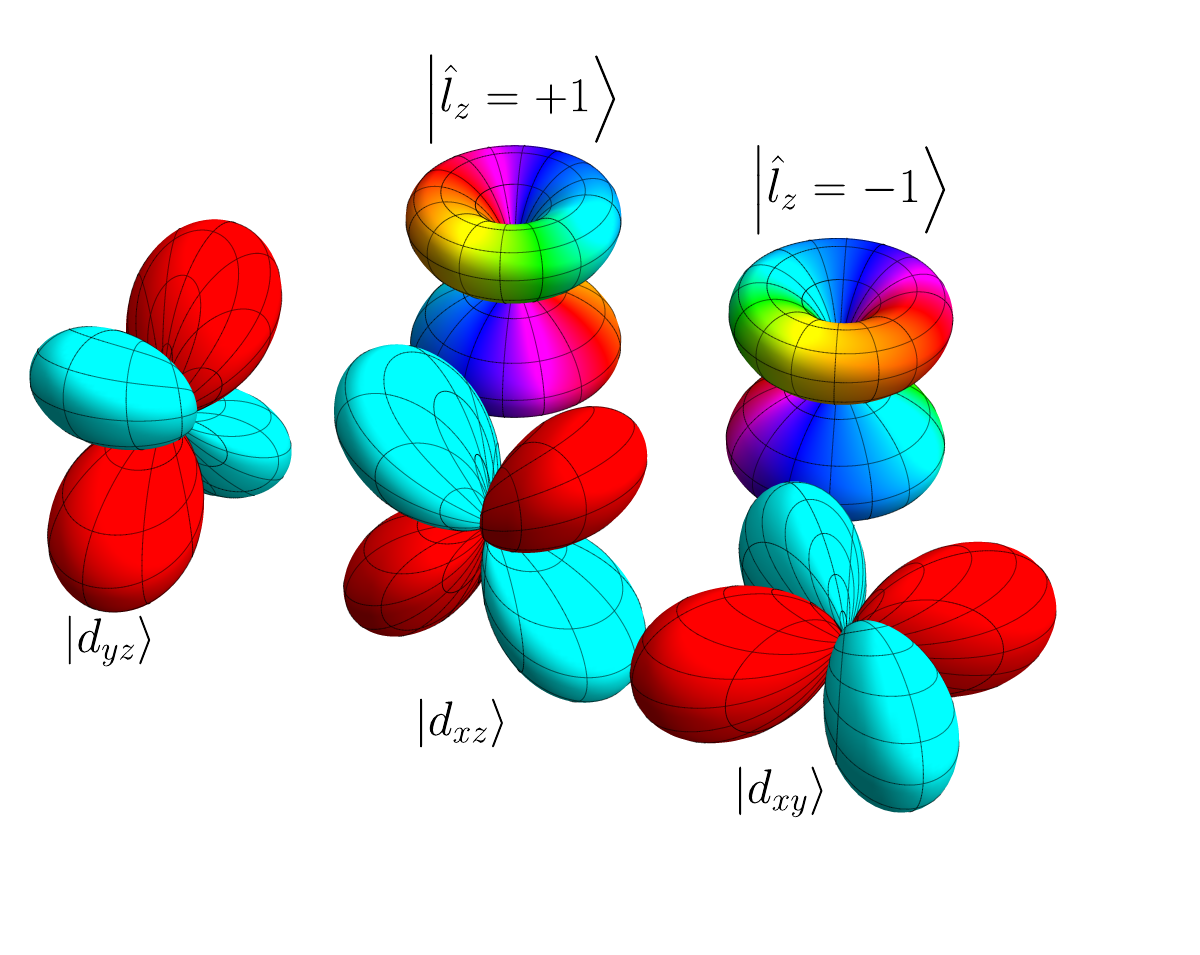}
	\end{center}
	\caption{Graphical representation of the real $\ket{d_{\alpha\beta}}$ and imaginary $\ket{\hat{l}_{z}=\pm1}$ orbital wavefunctions. The surface represents the absolute magnitude of the wavefunction whereas the color represents the phase.}
	\label{fig:Orb}
\end{figure}

Atomic orbitals of magnetic ions provide a link between the crystallographic structure and local magnetic moments and hence an avenue to couple structural and magnetic degrees of freedom.~\cite{Lee10:79,Khomskii22:11,Khomskii21:121,Streltsov20:10,Streltsov60:1121,Streltsov16:113,Tokura288:00}  A central parameter to controlling orbital order in materials is spin-orbit coupling which exists in magnetic ions with an inherit single-ion orbital degeneracy.  Given that spin-orbit coupling in multielectron atoms scales with the atomic number squared ($\lambda \propto Z^{2}$)\cite{Landau:book}, first row transition metals provide an opportunity to study magnetism when spin-orbit coupling is of a similar energy scale to the magnetic exchange and also energy scales of local structural distortions away from a perfect, for example, octahedral environment.  These energy scales have consequences to the nature of the real space magnetic orbitals which spatially order at low temperatures, breaking any degeneracy.  In particular is the situation of complex orbitals and real orbitals illustrated in Fig. \ref{fig:Orb}. In this paper we investigate the interplay between spin-orbital physics in MgV$_{2}$O$_{4}$ and apply neutron spectroscopy to obtain information on the underlying orbital order pointing to ordering of complex orbitals.

\begin{figure*}
	\begin{center}
		\includegraphics[width=185mm,trim=1.1cm 3.1cm 1.1cm 3.8cm,clip=true]{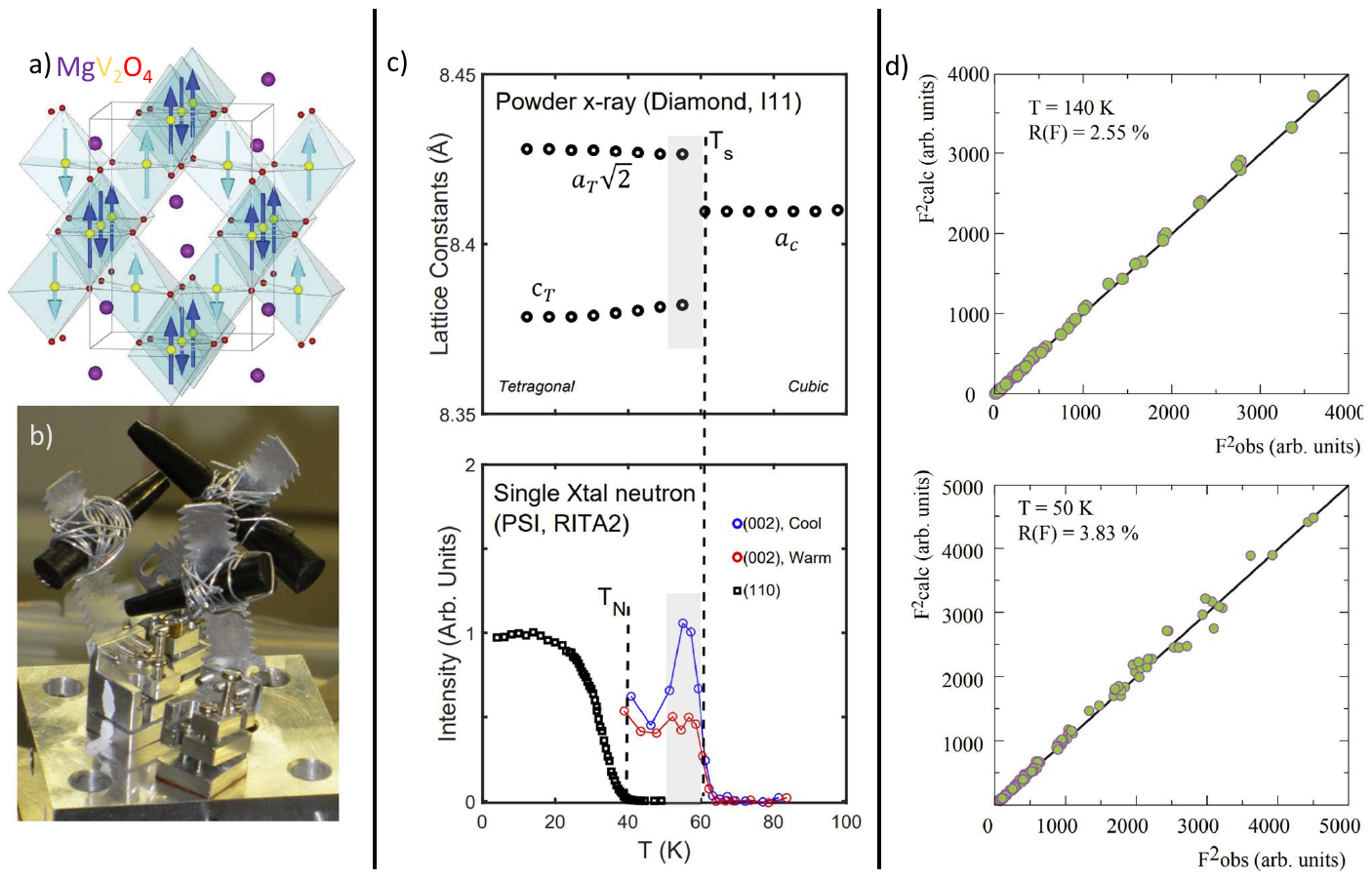}
	\end{center}
	\caption{$(a)$ The low temperature nuclear and magnetic structure of MgV$_{2}$O$_{4}$ obtained from our neutron diffraction results. $(b)$ Our coaligned single crystals.  $(c)$  Comparison of powder x-ray and neutron results illustrating the structural and magnetic transitions.  The shaded region indicates where both tetragonal and cubic phases coexist.  $(d)$ Single crystal neutron diffraction refinements of the nuclear structure in the high temperature T=140 K cubic phase and low temperature T=50 K tetragonal phases. The calculated Bragg peak structure factor squared ($F^{2}_{calc}$) is plotted against the measured structure factor squared ($F^{2}_{obs}$) as measured on D9 (ILL).}
	\label{fig:figure1}
\end{figure*}

Consisting of orbitally degenerate $d^{2}$ trivalent V$^{3+}$ ($S$=1, $l_{eff}$=1) decorating a pyrochlore~\cite{Gardner10:82} sublattice of corner-sharing tetrahedra (Fig. \ref{fig:figure1} $a$), the spinel vanadates $A$V$_{2}$O$_{4}$ with $A$ = divalent cation, provide a platform for the interplay of the underlying geometric constraints of a frustrated lattice~\cite{Moessner01:7} populated with strongly correlated electrons, each with charge, spin, and orbital degrees of freedom.~\cite{Lee10:79,Radaelli05:7}  The structural, electronic, and magnetic properties of the $A$V$_{2}$O$_{4}$ series are based on two broadly different classes with the $A$-site either being magnetic (as is the case for (Mn,Fe,Co)V$_{2}$O$_{4}$) or not (for example in (Mg,Zn)V$_{2}$O$_{4}$).~\cite{Tsurkan21:926}  These two classes of materials display some noteable differences with the size of the ordered moment on the V$^{3+}$ reportedly stronger when the $A$-site of the spinel structure is magnetic~\cite{Garlea08:100}. Also, magnetic $A$-site vanadate spinels display noncollinear magnetic order in contrast to their nonmagnetic counterparts.~\cite{Krishna19:100}  For simplicity towards investigating the orbital ground state of the V$^{3+}$ ion in this class of compounds, we discuss in this paper the case where the $A$-site is nonmagnetic. Given the orbital degeneracy on the V$^{3+}$ site, (Mg,Zn)V$_{2}$O$_{4}$ compounds exhibit a Jahn-Teller distortion with a contraction along the $c$-axis accompanying a cubic to tetragonal structural phase transition at $T_{S}$.  This distortion relieves the underlying geometric magnetic frustration and hence is followed by the formation of long-range collinear antiferromagnetic order at $T_{N}<T_{S}$.  The ordered moments in the N\'eel phase are reduced from expectations of a spin-only moment of $gS=2\ \mu_{B}$ with values of, for example, $\simeq$ 0.5 $\mu_{B}$ reported for MgV$_{2}$O$_{4}$.~\cite{Wheeler10:82}

The specific case of MgV$_{2}$O$_{4}$ exhibits an orbitally driven Jahn-Teller structural transition at T$_{S} \simeq$ 60 K and a N\'eel transition at T$_{N} \simeq$ 40 K.  The exact nature of the low temperature orbital order is ambiguous because two types of low-temperature orbital order have been proposed for this compound.  The first, termed Real Orbital Ordering (ROO), corresponds to orbital ordering of the real $t_{g}$ and $e_{g}$ orbitals which are eigenstates in the limit that the crystal field imposed on the V$^{3+}$ ion from the surrounding oxygen octahedra is large. This model has been advocated based on x-ray and electron beam diffraction data.~\cite{Niitaka13:111} An alternate suggestion~\cite{Tchernyshyov04:93} has been made for ordering of the complex basis of the $\hat{L}_{z}$ observable which are complex linear combinations of the real $d$-orbitals.  This is denoted as Complex Orbital Ordering (COO) and is the basis state used in the weak-intermediate crystal field limit for transition metal ions.  Based on powder neutron spectroscopy and diffraction, it has also been suggested that the orbital ordering could possibly be intermediate between the two ROO and COO extremes.~\cite{Wheeler10:82}  A graphical representation of the two different extremes is illustrated in Fig. \ref{fig:Orb}.

A theoretical study outlined in Ref. \onlinecite{Perkins07:76} suggested neutron spectroscopy as a means of distinguishing between COO and ROO orders at low temperatures.  In particular the study noted the importance of spin-orbit coupling with COO giving multiple magnetic branches in the neutron response and also a larger magnetic zone-center gap than would be expected in the ROO model.  Motivated by this, we investigate the magnetic neutron response in single crystals of MgV$_{2}$O$_{4}$.  We study the hysteretic critical dynamics using diffraction, measure the dynamic response, and develop a theory to describe the magnetic excitations from the spin-orbital ground state.  Given the spatially diffuse and correlated nature of the spin-orbital states used for our theory, we term this an excitonic approach.

We apply this excitonic approach to model our neutron spectroscopy results using the single-ion states and then treating the single-ion Hamiltonian (including spin-orbit coupling) and exchange energetics equally. \cite{Sarte19:100,Sarte20:102,Buyers75:11,Sarte20:32}  This model combined with the data favors the ordering of complex orbitals (COO) in MgV$_{2}$O$_{4}$ and illustrates the importance of using a complex orbital basis for understanding the properties of the first-row transition metal ions.

This paper is divided into four sections.  First, we outline the materials preparation methodology and experimental techniques used to probe the magnetic fluctuations and critical scattering.  Second, we outline the spectroscopic experimental results of both the low-energy spin-wave excitations and a higher energy optic-like mode.  Third, we present a theory for the magnetic excitations including spin-orbit coupling and the orbitally degenerate ground state.  Finally, we investigate the hysteretic effects resulting from the orbital degeneracy using energy integrated neutron diffuse scattering.

\section{Experimental Information}

\textit{Materials preparation:} Single crystals (Fig.\ \ref{fig:figure1} $b$) of MgV$_{2}$O$_{4}$ were grown using the floating zone technique with details provided in the Appendix.    Given the extreme sensitivity of the magnetic properties to stoichiometry and chemical order in MgV$_{2}$O$_{4}$~\cite{Islam12:85}, we have characterized our single crystals using both thermodynamic and scattering probes with neutrons and x-rays.  The diffraction results are summarized in Fig. \ref{fig:figure1} $(c)$, with high temperature structural (T$_{S}$ $\simeq$ 60 K) and magnetic (T$_{N}$ $\simeq$ 40 K from the magnetic $\vec{Q}$=(1,1,0) Bragg peak) transitions consistent with the published literature~\cite{Mamiya97:81,Wheeler10:82}.

\textit{Characterization:} Figure \ref{fig:figure1} $(c)$ illustrates powder diffraction measurements using synchrotron x-rays (Diamond, I11) and neutrons (PSI, RITA-II).  Powder synchrotron data shows a first order transitions at T$_{S}$ $\simeq$ 60 K from a high temperature cubic to a low temperature tetragonal phase.  This is confirmed on single crystal neutron diffraction data following the $\vec{Q}$=(0,0,2) Bragg peak on warming and cooling.  A large hysteresis in the neutron intensity is observed between warming and cooling over the same temperature range where synchrotron x-rays measure a coexistence of tetragonal and cubic phases.  This is further discussed in the Appendix where this region was found to extend from $\sim$ 55-60 K.  We have further confirmed that the structural properties of single crystals are consistent with the published structural phases through single crystal neutron diffraction data (D9, ILL).  The refinement results are summarized in Fig. \ref{fig:figure1} $(d)$ in both the cubic (T=140 K) and tetragonal (T=50 K) phases with the structural parameters listed in the Appendix.  Fig. \ref{fig:figure1} $(d)$ illustrates the calculated Bragg peak intensities ($\propto F^{2}$) as a function of the measured intensities with a straight line indicative that the models describe the results well.  The slope of the line a calibration factor scaling experiment and calculated values.  We have confirmed the magnetic structure to be consistent with that outlined in the literature for ZnV$_{2}$O$_{4}$~\cite{Reehuis03:35} using polarized neutrons (DNS, MLZ).  The results of this are schematically shown in Fig. \ref{fig:figure1} $(a)$ with the magnetic V$^{3+}$ magnetic moments pointing along the tetragonal $c$-axis.

\textit{Critical magnetic fluctuations:} In this paper we discuss energy-integrated polarized diffuse scattering measurements sensitive to the magnetic critical scattering and neutron inelastic data probing the spin-orbital dynamics.  To investigate the magnetic critical scattering as a function of temperature, we studied the polarized neutron cross section using an $xyz$ geometry provided by the DNS diffractometer (MLZ, Munich)~\cite{Su15:1}.  Further details of the polarized beam measurements and analysis are provided in the Appendix.

\textit{Neutron Spectroscopy:} To probe the magnetic dynamics sensitive to the spin-orbital ground state, neutron spectroscopy was performed on four different spectrometers.  To study the high-energy excitations which represent spin-orbital excitations we used the MAPS spectrometer. The sample was oriented so that $\vec{k}_{i}$ was oriented along the $c^{*}$ axis and the intensity along $L$ integrated as typically done with time of flight direct geometry spectrometers to measure one or two dimensional spin excitations. To obtain an overview of the low-energy and low dimensional dispersive dynamics, the multi rep rate option on MERLIN~\cite{Bewley06:385} was exploited combined with rotating the crystal.  Further data was taken on the EIGER triple-axis spectrometer~\cite{Stuhr16:853} to characterize weakly three dimensional dispersive excitations sensitive to interactions between chains. Finally, to probe the low-energy gapped excitations, sensitive to local single-ion anisotropy, we used the cold neutrons on the RITA-II spectrometer~\cite{Lefmann00:283}.

\section{Experimental results}
\label{Sect:Excitations}

\subsection{Low-energy dispersive dynamics}

\begin{figure}
	\begin{center}
		\includegraphics[width=90mm,trim=2cm 2cm 0.5cm 2cm,clip=true]{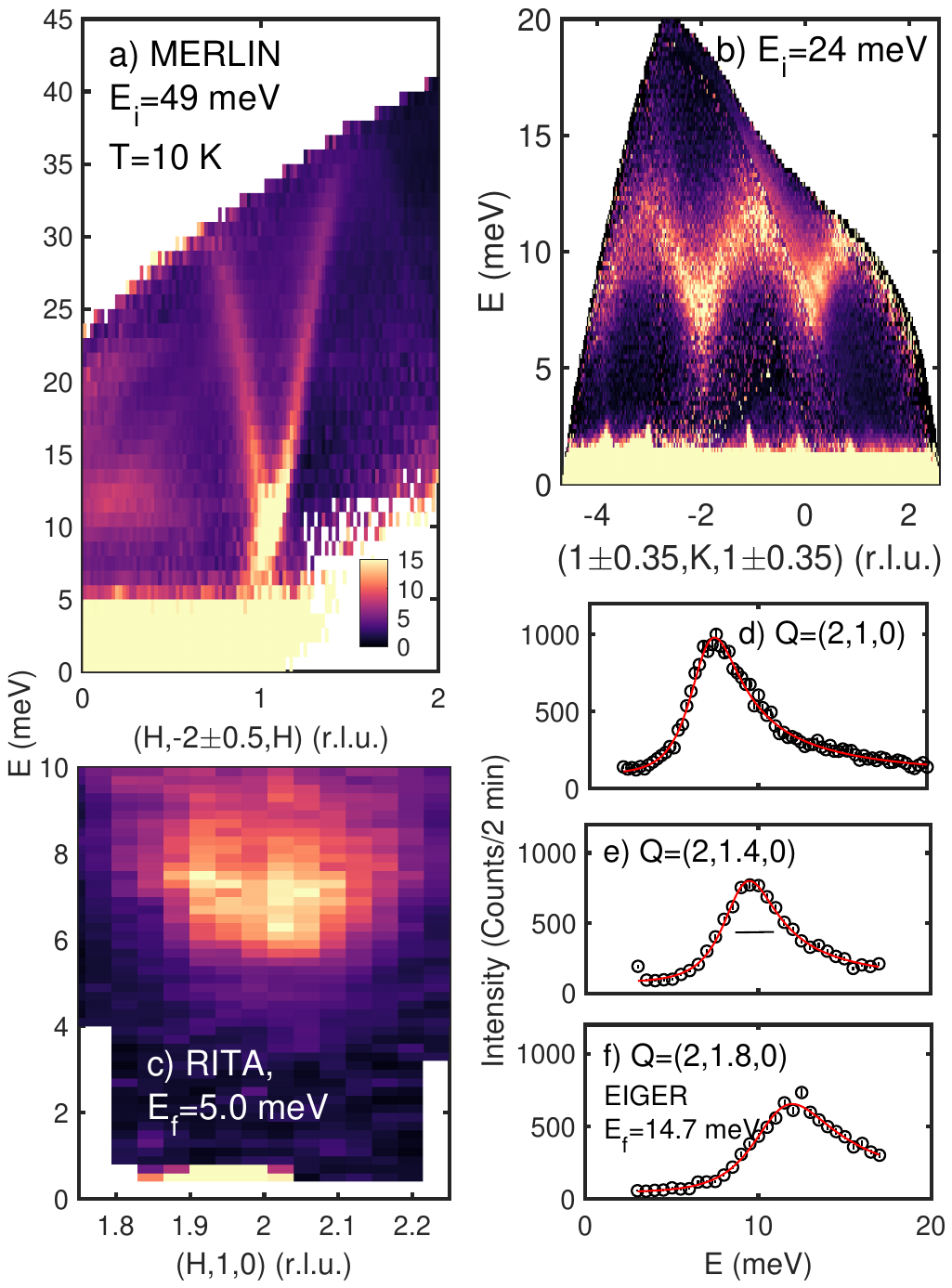}
	\end{center}
	\caption{$(a)$ illustrates the magnetic excitations, taken from MERLIN (E$_{i}$=49 meV) up to $\sim$ 40 meV showing a band of excitations. This cut is taken along the chain direction.  $(b)$ illustrates MERLIN (E$_{i}$=24 meV) displaying a comparatively weaker dispersion perpendicular to the chain direction.  $(c)$ displays high resolution data taken on RITA-II displaying an excitation gap reflecting underlying anisotropy.  $(d-f)$ shows data taken from EIGER characterizing the dispersion perpendicular to the chains.}
	\label{fig:dispersion_lowE}
\end{figure}

The magnetic excitations characterizing the underlying spin-orbital ground state below the N\'eel temperature T$_{N}$ are summarized in Fig. \ref{fig:dispersion_lowE}.  The overall dispersion is displayed in Fig. \ref{fig:dispersion_lowE} $(a)$ taken on the MERLIN spectrometer with E$_{i}$=49 meV.  This constant momentum slice illustrates a strong band of magnetic excitations which extends up to $\sim$ 35 meV.  Perpendicular to this direction, a much more weakly dispersive mode is illustrated in Fig. \ref{fig:dispersion_lowE} $(b)$.  This reflects the underlying strongly one dimensional coordination~\cite{Lee04:93} of the V$^{3+}$ ions.  The dispersion is further studied through a series of constant momentum scans using the EIGER thermal triple-axis spectrometer in Fig. \ref{fig:dispersion_lowE} $(d-f)$. Analyzing the low-energy fluctuations with the cold triple-axis spectrometer RITA-II in Fig. \ref{fig:dispersion_lowE} $(c)$, illustrates an energy gap of $\sim$ 6 meV in the fluctuation spectrum. These results demonstrate the presence of strongly one-dimensional magnetic fluctuations with an energetic gap due to crystalline anisotropy.  We discuss the origin of these below when we present an excitonic model for the spin-orbital excitations where we also discuss the relative magnetic exchange energies.

\begin{figure}
	\begin{center}
		\includegraphics[width=110mm,trim=4cm 2.5cm 0cm 1cm,clip=true]{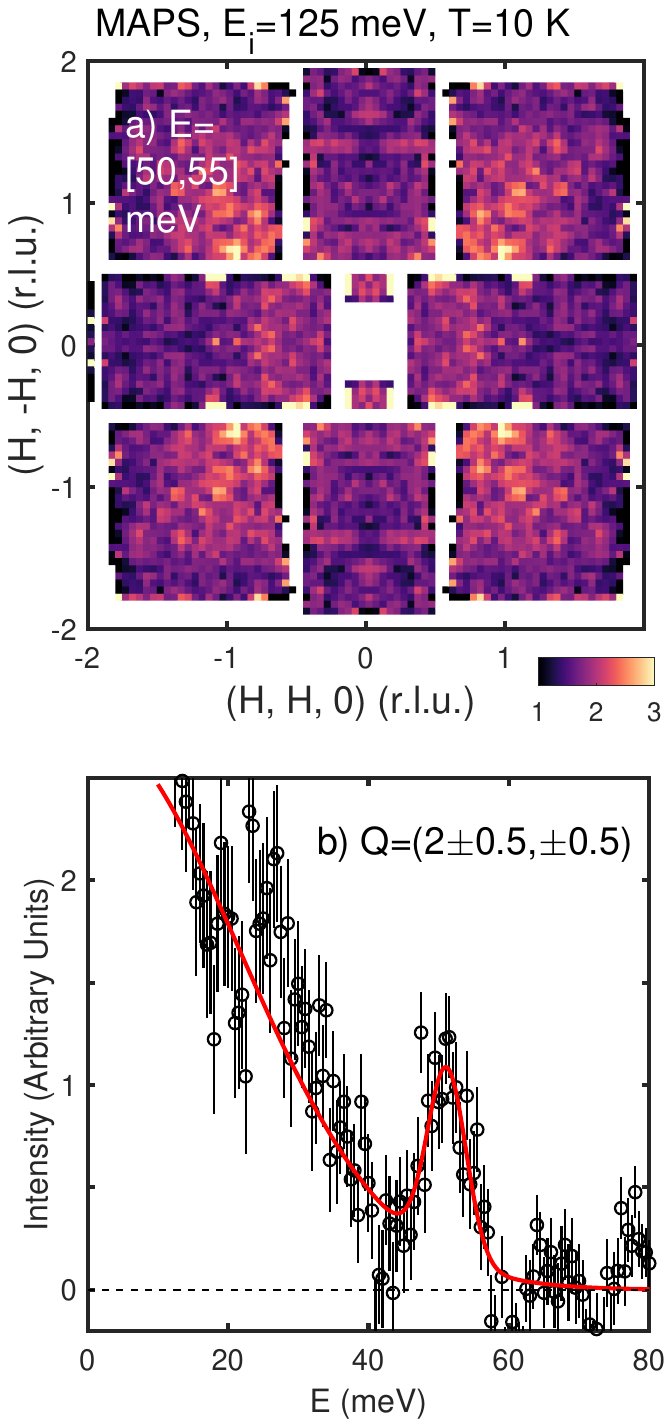}
	\end{center}
	\caption{$(a)$ displays a constant momentum slice at E=[50,55] meV taken using the MAPS spectrometer.  $(b)$ illustrates a background-corrected constant momentum cut showing a well-defined peak in energy indicative of a second higher energy magnetic band. The red line shows a fitted Gaussian on an exponentially decaying background peaked at 51 meV with a FWHM of 3 meV.}
	\label{fig:dispersion_highE}
\end{figure}

\subsection{Higher energy gapped mode}

One of the distinctions between COO and ROO orbital ordering scenarios is the presence of an optical-like excitation at higher energies~\cite{Perkins07:76} corresponding to a transition between the $j_{eff}=2$ and $j_{eff}=1$ manifolds (as theoretically outlined below).  A higher energy band of magnetic excitations is investigated using the MAPS spectrometer.  The results are summarized in Fig. \ref{fig:dispersion_highE}.  Fig. \ref{fig:dispersion_highE} $(a)$  displays a constant energy slice showing weakly correlated excitations near $\vec{Q}$=(2,0,0) and equivalent positions.  A background (taken from large momentum detectors) corrected constant momentum cut is shown in Fig. \ref{fig:dispersion_highE} $(b)$ displaying a well-defined peak at $\sim$ 50 meV.  We discuss the origin of this additional excitation and connect it with the low-energy response below by investigating the spin-orbital neutron response theoretically.

\begin{figure*}
	\begin{center}
		\includegraphics[width=160mm,trim=0cm 0cm 0cm 0cm,clip=true]{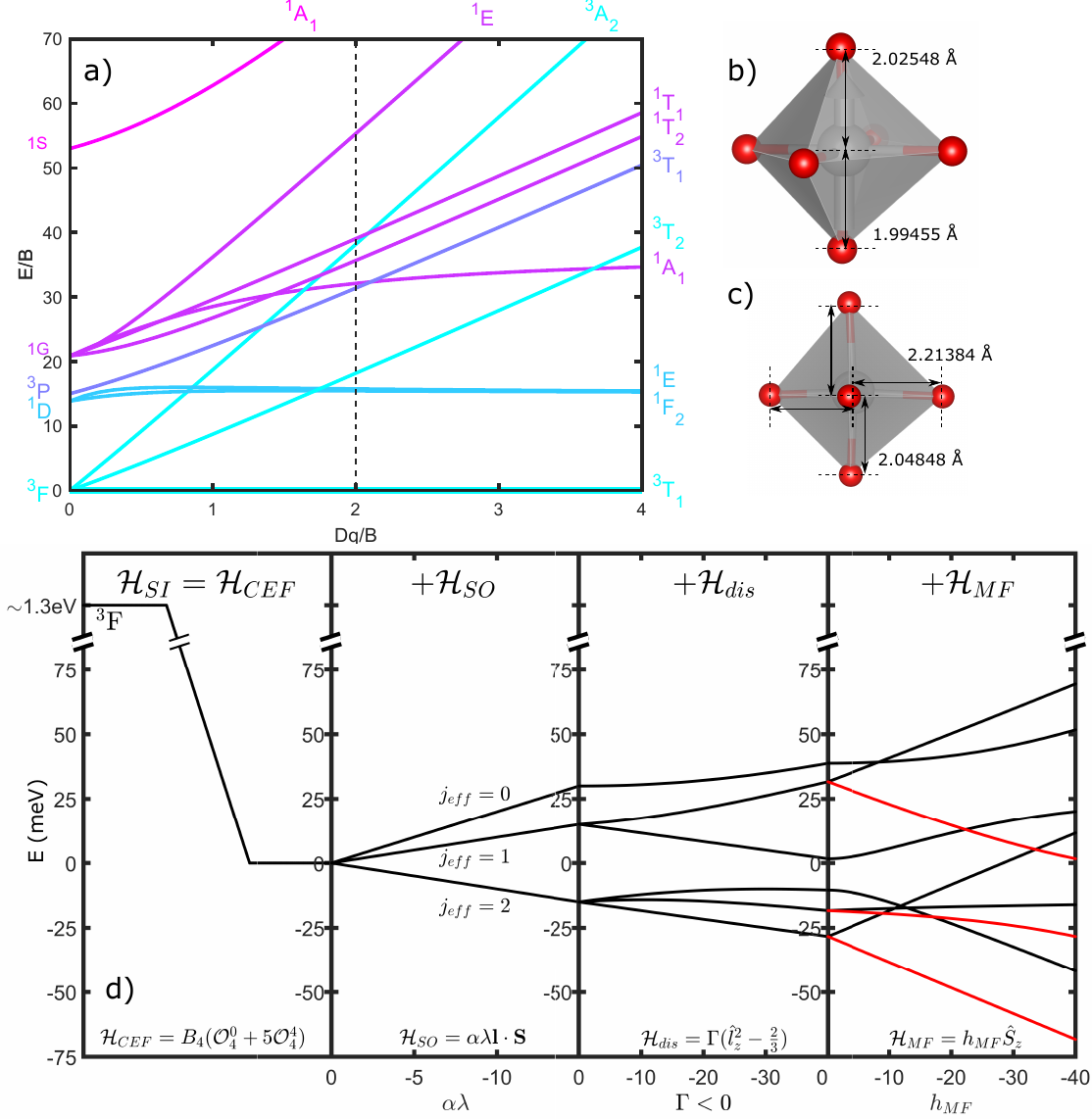}
	\end{center}
	\caption{$(a)$ Tanabe-Sugano diagram for a $d^{2}$ ion in an octahedral crystal field. The Racah parameters have been chosen such that $C/B= 4.43$~\cite{McClure:book}. For the discussion of the crystal field parameters, we have followed Ref. \onlinecite{Abragam:book} (Table 7.3) and used B=0.11 eV and $Dq$=0.22 eV, indicated by the dashed line. The VO$_{6}$ octahedron as viewed from $(b)$ an isometric viewpoint and $(c)$ the $c$-axis. The octahedron is tetragonally distorted, as evidenced by the shorted bond lengths along $c$. A further trigonal distortion is present along [111] leading to two different bond lengths in the $a$-$b$ plane. $(d)$ Eigenvalues of the single-ion Hamiltonian for an octahedrally coordinated V$^{3+}$ ion ($S=1$, $l=1$) subject to a crystallographic distortion as described in Sect. \ref{Sect:Single-ion physics}. Red lines indicate the ground state and excitated states for which dipole-allowed transitions exist.}
	\label{fig:Theory_figure_1}
\end{figure*}

\section{Theory}

In MgV$_{2}$O$_{4}$ the magnetic V$^{3+}$ (3$d^{2}$) ions form a pyrochlore lattice with the V$^{3+}$ ions on the spinel B sites surrounded by an octahedral coordination of oxygen O$^{2-}$ (Fig. \ref{fig:Theory_figure_1} $b,c$) which determines the orbital ground state. Neighboring V$^{3+}$ ions occupy edge-sharing octahedra with non-magnetic tetrahedrally-coordinated Mg filling the voids between VO$_{6}$ octahedra. The magnetic inter-ion interactions are likely governed by a combination of direct $d$-$d$  orbital overlap and oxygen mediated superexchange via the $\sim 90^{\circ}$ V-O-V bonds. 

In the rare earth pyrochlores~\cite{Rau19:100}, where the B site is occupied by the heavier $4f$ ions, spin-orbit coupling ($\lambda \sim Z^{2}$)~\cite{Landau:book} is the dominant energy scale and motivates the projection onto a ground state manifold $J\equiv L+S$.  The crystalline electric field further splits the ground state with the resulting energy distribution dependent on the species of ion and local environment. In the case that the ground state is a Kramers doublet, a projection onto an effective $S=1/2$ pseudospin~\cite{Ross11:1,Savary12:109,Thompson17:119} has been utilized and provided the motivation for seeking out quantum spin liquid~\cite{Savary16:80} behavior in these materials.

The physics of the $3d$ pyrochlores is unlike that of their rare earth cousins owing to a different hierarchy of single-ion energy scales. Particularly, in $3d$ ions, the spin-orbit coupling ($\lambda$) is smaller than the crystalline electric field ($Dq$)~\cite{Abragam:book,Sarte18:98,Sarte18:98_2}. With typical energy scales of $Dq$ $\sim$ 0.1 eV and $\lambda$ $\sim$ 10 meV, spin-orbit coupling is a perturbation on the crystalline electric field Hamiltonian and the ground state is defined by the orbital angular momentum $L$. Within this ground state manifold, the spin-orbit coupling, magnetic exchange and Jahn-Teller energy scales are comparable. In the $3d$ transition metal compounds it is therefore necessary to consider both single-ion spin-orbital energy scales ($\mathcal{H}_{SI}$) and the corresponding magnetic exchange interactions ($\mathcal{H}_{exch}$)\cite{Sarte18:98}

\begin{equation*}
	\mathcal{H}=\mathcal{H}_{SI}+\mathcal{H}_{exch}.
\end{equation*}

\noindent We will begin by discussing the single ion physics of the $3d^{2}$ ions to understand the magnetic ground state before considering the magnetic inter-ion exchange in vanadium spinels. Using a Green's function formalism, the dynamical structure factor measured with neutrons will then be calculated using the random phase approximation (RPA) applied to MgV$_{2}$O$_{4}$, treating the single-ion Hamiltonian which produces a quantized multilevel ground state explicitly. 

\subsection{$\mathcal{H}_{SI}$ - Single-ion physics}
\label{Sect:Single-ion physics}

The single ion physics discussed in this section is schematically shown in Fig. \ref{fig:Theory_figure_1} which determines the quantized spin-orbital ground state of the magnetic V$^{3+}$ ions.  We consider first the free-ion V$^{3+}$ ground state followed by the effects of the crystalline electric field, spin-orbit coupling, distortions from an octahedral environment, and finally the Zeeman-like molecular field originating from magnetic ordering in the low temperature N\'eel phase.

\subsubsection{$^{3}F$ - Free ion ground state}

For the case of a free V$^{3+}$ ions with 2 electrons in the five degenerate $d$ orbitals, the ground state is determined by Hund's rules which defines $L=3$ and $S=1$. This fixes the orbital ground state of the free-ion V$^{3+}$ state to be $^{3}F$ in spectroscopic notation.
 
\subsubsection{$\mathcal{H}_{CEF}$ - Crystalline electric field}
 
As discussed above, the dominant energy scale for magnetic $3d$ is the crystalline electric field originating from the O$^{2-}$ ions forming an octahedron around the V$^{3+}$ ion.   Application of an octahedral crystalline electric field in terms of Steven's operators $O_{l}^{m}$ gives the following

\begin{equation*}
	\mathcal{H}_{CEF}=B_{4}\left( O_{4}^{0}+5O_{4}^{4}\right)
\end{equation*}

\noindent where $B_{4}<0$ for $d^{2}$ ions~\cite{Abragam:book}. The energy spectrum resulting from this crystalline electric field Hamiltonian is schematically displayed in Fig. \ref{fig:Theory_figure_1} $(a)$ (Tanabe-Sugano diagram) and parameterized in terms of the crystal field strength ($Dq$) and the Racah parameter $B$ which physically corresponds to the energy cost associated with the Coulomb repulsion.  The limit $Dq/B \rightarrow 0$ corresponds to the weak crystal field limit where Hunds rules across all $d$-orbitals applies.  In the large $Dq/B$ limit, the crystal field energy scale is dominant and in some transition metal ions (such as Co$^{2+}$) can result in spin transitions.

Figure \ref{fig:Theory_figure_1} $(a)$  shows that the octahedral crystal field splits the orbital $^{3}$F ground state into two triplets ($^{3}T_{1,2}$) and a singlet ($^{3}A_{2}$), with the $^{3}T_{1}$ triplet being the orbital ground state. The splitting between $^{3}T_{1}$ and the first excited triplet is $8Dq = 480B_{4} \approx $ 1.8 eV~\cite{Abragam:book} (in the limit of small $Dq$).  Each of the multiplets under the octahedral crystal field forms an irreducible representation of the octahedral double group, $O'$~\cite{Tinkham:book}. By the orthogonality of different irreps of the same group, for an operator that is invariant under all octahedral symmetry operations, matrix elements between multiplets are zero~\cite{Tinkham:book}. Assuming further symmetry lowering terms away from an octahedral field are small and the large energy separation in comparison to the temperatures of interest, one can justifiably work in the ground state $^{3}T_{1}$ multiplet, neglecting the excited states. 

Considering energy levels beyond the $^{3}F$ manifold in Fig. \ref{fig:Theory_figure_1} $(a)$, the octahedral field mixes in the excited $^{3}P$ state (Fig. \ref{fig:Theory_figure_1} $a$ when $Dq/B \rightarrow 0$).  The lowest energy orbital state of the $^{3}P$ manifold has the same symmetry being a $^{3}T_{1}$ triplet~\cite{Abragam:book} and therefore the overall groundstate is a linear combination of the two triplet states from the $^{3}F$ and $^{3}P$ manifolds 

\begin{equation}
	\ket{\psi(^{3}T_{1})} = \epsilon \ket{^{3}F(^{3}T_{1})}+\tau \ket{^{3}P(^{3}T_{1})}   
	\label{Gamma4Weak:eq}
\end{equation}

\noindent where $\epsilon^{2}+\tau^{2} = 1$~\cite{Abragam:book}. One can represent the ground state orbital triplet using the fictitious orbital angular momentum operator, $\mathbf{\hat{l}}$ via the transformation $\mathbf{\hat{L}}=\alpha \mathbf{\hat{l}}$, where the projection factor $\alpha<0$ can be read off the block describing the ground state manifold after projecting the matrices $\hat{L}_{\alpha}$ onto the space spanned by the eigenvectors of $\mathcal{H}_{CEF}$. For the $\ket{^{3}F(^{3}T_{1})}$ block, $\alpha=-\frac{3}{2}$~\cite{Lane21:104}, whilst for the $\ket{^{3}P(^{3}T_{1})}$ state, $\alpha = +1$. The admixture of these two states leads to an effective projection factor that varies between these two extremal values, $\alpha = \frac{5}{2}\tau^{2}-\frac{3}{2}$. 

It is instructive to discuss the orbital triplet ground state in terms of the electronic orbital basis states which have the advantage over the eigenstates of the observables $L^{2}$ and $L_{z}$ of being real and therefore give a greater connection to the microscopic picture of electron hopping through chemical bonding. In the $3d$ ions, the outer electrons partially fill $d$-orbitals, which are split in an octahedral crystal field into the triply degenerate $t_{2g}$ level ($d_{xy},d_{yz},d_{xz}$) and doubly degenerate $e_{g}$ level ($d_{x^2-y^2},d_{3z^{2}-r^{2}}$). The energy splitting between $t_{2g}$ and $e_{g}$ levels is set by $10 Dq$, and therefore this parameter is key in distinguishing the weak (small $Dq$) from the strong (large $Dq$) crystal field limits.  For weak crystal fields it is more natural to use a complex orbital basis that are eigenstates of the observables $L^{2}$ and $L_{z}$.  In the large crystal field limit, the real $t_{2g}$ and $e_{g}$ states are the convenient basis. 

Neglecting the spin-orbit coupling and further distortions, the ground state for 2 electrons in the $3d$ orbitals is triply degenerate with each member of ground state manifold having two of the $t_{2g}$ levels occupied. In terms of these real basis states one can represent the ground state triplet with the admixture caused by the crystal field (Eq. \ref{Gamma4Weak:eq}) equivalently as

\begin{equation}
	\ket{\psi(^{3}T_{1})}=\mathrm{cos}\theta\ket{t_{2g},t_{2g}}-\mathrm{sin}\theta\ket{t_{2g},e_{g}}.
	\label{Gamma4Strong:eq}
\end{equation}

\noindent The first of these two basis states has two occupied $t_{2g}$ levels as expected in an octahedral crystal field, and the second has an occupied $e_{g}$ level but has the same symmetry as $\ket{t_{2g},t_{2g}}$. By diagonalizing the energy matrix for a $d^{2}$ ion with both a Coulomb term and an octahedral crystal field~\cite{Abragam:book} (Fig. \ref{fig:Theory_figure_1}$a$) one finds $\mathrm{tan}2\theta=12B/(9B+10Dq)$ and using the correspondence between the two descriptions (Eqs. \ref{Gamma4Weak:eq} and \ref{Gamma4Strong:eq}) one can quantify the fraction of the $^{3}P$ level in the ground state orbital triplet, $\tau=\frac{1}{\sqrt{5}}(\mathrm{cos}\theta-2\mathrm{sin}\theta)$. With $B=0.11$ eV and $Dq=0.22$ eV, as expected for a free V$^{3+}$ ion, we have $\tau\approx0.27$. We therefore expect a projection factor of $\alpha\approx-1.32$ in comparison to a value of -1.5 in the absence of mixing.  We discuss this parameter later in context of the reported ordered magnetic moment and the possibility for quantum fluctuations. 

\subsubsection{$\mathcal{H}_{SO}$ - Spin-orbit coupling}

The next largest term (illustrated in Fig. \ref{fig:Theory_figure_1} $d$) in the single-ion Hamiltonian for $3d$ ions is the spin-orbit coupling

\begin{equation*}
	\mathcal{H}_{SO}=\lambda \mathbf{\hat{L}}\cdot \mathbf{\hat{S}}=\alpha \lambda \mathbf{\hat{l}}\cdot \mathbf{\hat{S}}
\end{equation*}

\noindent where $\lambda>0$ for $d^{2}$ systems\cite{Yosida:book}. This term splits the ground state ninefold spin-orbital manifold into three $\mathbf{\hat{j}}=\mathbf{\hat{l}}+\mathbf{\hat{S}}$ levels according to the Land{\'e} interval rule with $j_{eff}$=0, 1, 2 (Fig. \ref{fig:Theory_figure_1} $d$). In $d^{2}$ ions, the ground state is the $j_{eff}=2$, separated from the $j_{eff}=1$ state by $|2\alpha \lambda|$~\cite{Abragam:book}. In the language of the strong crystal field $t_{2g}$ levels, the effect of the inclusion of spin-orbit coupling is to mix the $d_{xz}$ and $d_{yz}$ orbitals in complex combinations such that three basis states in which the Hamiltonian is diagonal become $\ket{l_{z}=0}=\ket{d_{xy}}$ and $\ket{l_{z}=\pm 1}=\frac{1}{\sqrt{2}}(\ket{d_{xz}}\pm i\ket{d_{yz}})$~\cite{Khomskii:book}.

\subsubsection{$\mathcal{H}_{dis}$-Distortion}

In MgV$_{2}$O$_{4}$ at the low temperatures of interest here, the symmetry of the local VO$_{6}$ octahedra is not $O_{h}$, since a Jahn-Teller distortion, originating from the orbital degeneracy of the $d^{2}$ vanadium ion, occurs on the transition from the high temperature cubic phase to the low temperature tetragonal phase~\cite{Wheeler10:82}. This distortion results in the octahedra being subtly compressed along the fourfold $\hat{z}$-axis (illustrated in Fig. \ref{fig:Theory_figure_1} $b$).  The distortion can be modeled by the following term in the Hamiltonian 

\begin{equation}
	\mathcal{H}_{dis}=\Gamma \left(\hat{l}_{z}^{2}-\frac{2}{3}\right)
	\label{Hdis:eq}
\end{equation}

\noindent where $\Gamma <0$ for a compression. As displayed in Fig. \ref{fig:Theory_figure_1} $(d)$ the effect of this term is to break the five fold orbital groundstate $j_{eff}=2$ degeneracy into three levels with a groundstate doublet, an excited state doublet, and an excited state singlet.

Considering the effect of this distortion only on orbitals (without considering spin or spin-orbit coupling and in the limit of $|\Gamma| \gg |\lambda|$) and in terms of the strong crystal field basis of real orbitals, the effect of this axial distortion is to break the ground state $t_{2g}$ orbital triplet degeneracy, yielding a ground state orbital doublet and excited singlet. The distortion lowers the energy of the $d_{xy}$ orbital relative to the $d_{xz}$ and $d_{yz}$ orbital.  If we populate these two levels with two $d$ electrons applying Pauli's exclusion principle,  this results in a doubly degenerate ground state with a hole in either the $d_{xz}$ or $d_{yz}$ orbital.

We note that in addition to the primary tetragonal compression driven by orbital degeneracy, the VO$_{6}$ octahedra are trigonal distorted (Fig. \ref{fig:Theory_figure_1} $c$) -- a compression along the threefold $[111]$ axis (even within the high temperature cubic phase). Within the manifold of the groundstate $t_{2g}$ orbitals, the multiplet structure under this distortion is the same as the tetragonal distortion (Eq. \ref{Hdis:eq})~\cite{Khomskii:book}. In this paper, we will take advantage of the projection onto the ground state triplet, $\mathbf{\hat{L}}=\alpha \mathbf{\hat{l}}$ and treat the mixing of the $e_{g}$ levels as a small perturbation and so we can collect all contributions to the distortion Hamiltonian into a single distortion parameter, $\Gamma$. Under a dominant trigonal distortion,  the orbital spectrum has the form $\ket{a_{1g}}=\frac{1}{\sqrt{3}}(\ket{d_{xy}}+\ket{d_{xz}}+\ket{d_{yz}})$ with the excited doublet $\ket{e_{g}^{\pi}}=\pm\frac{1}{\sqrt{3}}(\ket{d_{xy}}+e^{\pm 2\pi i /3}\ket{d_{xz}}+e^{\mp 2\pi i /3}\ket{d_{yz}})$~\cite{Khomskii:book}. From crystallographic considerations, the low temperature tetragonal distortion discussed above is expected to be dominant, however one might expect a small difference in orbital coefficients arising due to the subleading trigonal distortion. Such a situation maybe supported by the small $\sim 8^{\circ}$ canting of the spin towards the apical oxygen reported in some diffraction studies~\cite{Wheeler10:82}. 

The orbital state for each electron can be written in general as

\begin{equation}
	\ket{\psi}=\alpha \ket{d_{xy}}+\beta e^{i\theta}\ket{d_{xz}}+\gamma e^{i\phi}\ket{d_{yz}}
	\label{OO:eq}
\end{equation}

\noindent where $\alpha^{2}+\beta^{2}+\gamma^{2}=1$ ensures normalization~. Based on the single-ion physics of the V$^{3+}$ ion ($3d^{2}$) alone, the orbital order in MgV$_{2}$O$_{4}$ is expected to be intermediate between the regimes of validity of real orbital order (ROO), which occurs when the tetragonal distortion dominates and complex orbital order (COO) where the spin-orbit coupling dominates, since the energy scales of the distortion and the spin-orbit coupling in the $3d$ ions are typically nearly comparable~\cite{Lane21:104,Sarte19:100,Cowley73:6}. The weightings of the $d_{\alpha\beta}$ orbitals for each electron within each regime are summarized in Table \ref{OO:Table}. We will discuss the nature of the orbital order further later in the paper based on the insight gained through our analysis.

\begin{table}[h]
	\caption{\label{OO:Table} Orbital wavefunction coefficients in common orbital ordering regimes for electron I, electron II. Coefficients correspond to Eq. \ref{OO:eq}. Column two states the dominant energy scale in each case. }
	\begin{ruledtabular}
		\begin{tabular}{ccccccc}
			OO & Dom. & $\alpha$ & $\beta$ & $\gamma$ & $\theta$ & $\phi$ \\ 
			\hline
			COO & $\lambda$ & 1,0 &0,$\frac{1}{\sqrt{2}}$ &0,$\frac{1}{\sqrt{2}}$ & 0,0 & 0,$\frac{\pi}{2}$ \\ 
			ROO & $\Gamma_{[001]}>0$& 1,0& 0,$\frac{1}{\sqrt{2}}$ &0,$\frac{1}{\sqrt{2}}$ & 0,0& 0,0 \\
			Trig. COO& $\Gamma_{[111]}$ &$\frac{1}{\sqrt{3}}$,$\frac{1}{\sqrt{3}}$ & $\frac{1}{\sqrt{3}}$,$\frac{1}{\sqrt{3}}$& $\frac{1}{\sqrt{3}}$,$\frac{1}{\sqrt{3}}$&0,$\frac{2\pi}{3}$ & 0,$\frac{-2\pi}{3}$ 
		\end{tabular}
	\end{ruledtabular}
\end{table}  

\subsubsection{$\mathcal{H}_{MF}$ - Molecular field below T$_{N}$}

The final term present in the single-ion Hamiltonian is the molecular field. This results from a mean field decoupling of the exchange interaction between coupled ions, and is required such that the single ion ground state about which one expands is correct. This term breaks time reversal symmetry 

\begin{equation*}
	\mathcal{H}_{MF}=h_{MF}\hat{S}_{z}
\end{equation*}

\noindent as is consistent with the establishment of long-range magnetic order. The magnitude of $h_{MF}$ will be discussed further once the inter-ion spin Hamiltonian, $\mathcal{H}_{exch}$, has been introduced. 

In terms of the single-ion Hamiltonian, the ground state order is expected to be $(S=1, l=+1)$ with positive $+l$ (instead of $-l$) selected due to the negative spin-orbit coupling constant in $d^{2}$ ions which promotes alignment of the spin and orbital moments.  This expected ground state gives rise to a magnetic moment $\mu=\mu_{B}(\mathbf{\hat{L}}+2\mathbf{\hat{S}})=\mu_{B}(-1.32+2)=0.68\mu_{B}$ which agrees well with the observed (reduced) magnetic moment, $\mu=0.47 \mu_{B}$.~\cite{Wheeler10:82} The discrepancy between the observed and calculated magnetic moment suggests the presence of quantum fluctuations, $S-\Delta S\approx 0.9$, as is to be expected given the small value of $S$ and reduced dimensionality originating from the frustrated lattice. This reduction in the ordered moment in turn contributes to a reduction in the molecular mean field over what would be expected for the full spin value.

\subsection{Inter-ion coupling}
\label{Sect:Inter-ion coupling}

We now discuss the inter-ion interactions present in MgV$_{2}$O$_{4}$. In the high temperature cubic phase, the V$^{3+}$ ions lie on an ideal pyrochlore lattice, with each nearest neighbor V-V bond equivalent by symmetry. In the case of antiferromagnetic nearest-neighbor interactions, the spin arrangement on a pyrochlore lattice is geometrically frustrated~\cite{Bramwell_98:10}, opening the possibility of spin liquid states~\cite{Moessner98:80,Savary16:80} resulting from the large ground state degeneracy. Pyrochlore systems with $S=1$ have been studied both theoretically~\cite{Iqbal19:9,Zhang19:122} and experimentally~\cite{Gardner99:83,Krizan15:92,Plumb19:15}, particularly in search of an understanding of the crossover from quantum to classical spin liquid behavior~\cite{Plumb19:15,Zhang19:122}. The vanadium spinels have also attracted great interest~\cite{Tchernyshyov04:93,DiMatteo05:72,Perkins07:76,Motome05:74,Chern10:81} since the orbital degree of freedom allows for Jahn-Teller distortions which break the ground state degeneracy and permit magnetic order. This behavior is observed in MgV$_{2}$O$_{4}$ at T$_{S}$ $\simeq$ 60 K, where the system becomes tetragonal before ordering antiferromagnetically at T$_{N} \simeq$ 40 K~\cite{Wheeler10:82}. The cubic to tetragonal structural transition renders the nearest neighbor bonds along $[HH0]$ and $[H0H]$ inequivalent.

\begin{figure}
	\begin{center}
		\includegraphics[width=\linewidth]{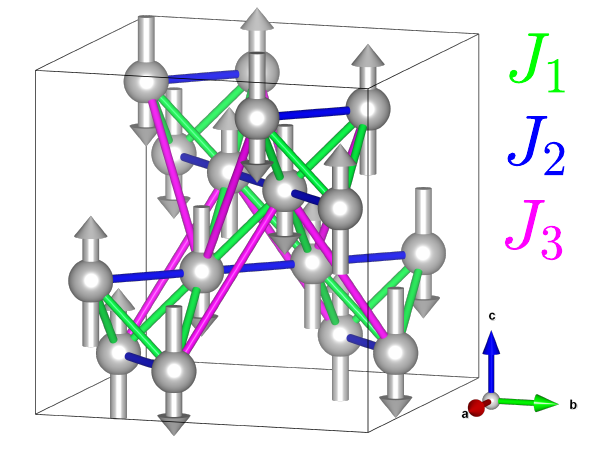}
	\end{center}
	\caption{Magnetic structure of MgV$_{2}$O$_{4}$ as viewed from an isometric viewpoint. The structure comprises chains in the $a$-$b$ plane with the nearest-neighbor bond, $J_{2}$. Inter-chain bond $J_{1}$ is frustrated and couples chains along $[H0H]$ and $[0HH]$. A longer range bond, $J_{3}$ couples perpendicular chains.  }
	\label{fig:Theory_figure_2}
\end{figure}

The orbital order plays an important role in the magnetic inter-ion exchange since both the $d$-$d$ and $d$-$p$ orbital overlap are dependent on the V$^{3+}$ orbital ground state, and ultimately where the electron hole lies. This provides a mechanism for a breaking of the equivalence of magnetic bonds both in terms of the amplitude of the direct and super-exchange interactions ~\cite{Khomskii:book}. The question of the magnetic exchange in the vanadium spinels has been discussed at length by Di Matteo \textit{et al}~\cite{DiMatteo05:72} and Perkins \textit{et al}~\cite{Perkins07:76}. We will now briefly summarize the findings of the aforementioned references~\cite{DiMatteo05:72,Perkins07:76} insofar as they are relevant to the magnetic exchange in MgV$_{2}$O$_{4}$.

Di Matteo \textit{et al} considered the limit where the tetragonal distortion is dominant (ROO - real electronic orbitals discussed above) and where the spin-orbit coupling is dominant (COO - complex basis described above when the crystalline electric field $Dq$ is weak)~\cite{DiMatteo05:72}. Both limits give rise to multiple magnetic and orbital configurations dependent on the underlying physical parameters~\cite{DiMatteo05:72} and described by the general Heisenberg spin Hamiltonian

\begin{equation*}
	\mathcal{H}_{exch}=\frac{1}{2}\sum_{ij}\mathcal{J}_{ij}\mathbf{\hat{S}}_{i}\cdot \mathbf{\hat{S}}_{j},
\end{equation*}

\noindent where the factor of $\frac{1}{2}$ accounts for double counting. In the case of ROO, only one possible orbital ground state is consistent with the experimentally observed magnetic structure~\cite{Wheeler10:82}. This ROO ground state comprises strong antiferromagnetic (AFM) bonds in the $xy$ plane and a weaker ferromagnetic coupling along $[H0H]$ and $[0HH]$. The ROO model thus suggests that MgV$_{2}$O$_{4}$ is comprised of strongly-coupled AFM chains in the $xy$ plane with weak FM coupling between chains. In each tetrahedron, two FM bonds are satisfied and two FM bonds are frustrated. 

In contrast, when the spin-orbit coupling is dominant (COO), there are two ground states consistent with the observed magnetic structure. Both possible orbital ground states have strongly coupled AFM chains in the $xy$ plane, however one of these states has two weak AFM and two weak FM bonds per tetrahedron -- thus no bond is frustrated. In the other possible COO ground state, the inter-chain bonds are weak AFM bonds, two of which are frustrated. Based on the parameters obtained from spectroscopy data for vanadium perovskites~\cite{Tsunetsugu03:68,Mizokawa96:54} and typical strengths of the spin-orbit coupling for V$^{3+}$ reported in the literature~\cite{Tchernyshyov04:93,Abragam:book}, Di Matteo \textit{et al.} concluded that the latter of the two COO states is likely the ground state~\cite{DiMatteo05:72}.

Finally we note that the predicted magnetic ground states in the ROO and COO limits differ only in the sign of the inter-chain coupling, with both inter-chain couplings of a similar strength~\cite{DiMatteo05:72}. It is interesting to speculate as to whether these two limits are continuously connected via a phase with zero inter-chain coupling and hence might be tuned through with the application of strain, using the tetragonal distortion, $\Gamma$, as a tuning parameter.  It is also interesting to note that a metal-insulator has been predicted in MgV$_{2}$O$_{4}$ based on ab initio calculations.~\cite{Baldomir08:403} 

\subsection{Neutron scattering intensity calculation}

\begin{table}[h]
\caption{\label{Table:indices} Summary of labeling convention for indices.}
\begin{ruledtabular}
\begin{tabular}{cc}
Index & Description \\ 
\hline
$\gamma$, $\gamma'$ & V$^{3+}$ sites   \\ 
$p$, $q$ & spin-orbital crystal field states  \\ 
$\alpha$, $\beta$, $\mu$, $\nu$ & Cartesian coordinates  \\ 
\end{tabular}
\end{ruledtabular}
\end{table}

In order to calculate the neutron scattering response, we employ a spin-orbital exciton model in terms of Green's functions as applied previously to describe multi-level systems \cite{Buyers75:11,Sarte19:100,Sarte20:102,Lane21:104,Lane21:104_2,Lane:preprint}. A full derivation for a general single-$\mathbf{Q}$ multi-level magnetic system is presented in Ref. \onlinecite{Lane:preprint}, here we quote only the key results. The neutron scattering intensity is proportional to the structure factor

\begin{equation*}
	S({\bf{q}},\omega)=g_{L}^{2}f^2({\bf{q}})\sum_{\alpha \beta} (\delta_{\alpha \beta}-\hat{q}_{\alpha}\hat{q}_{\beta}) S^{\alpha \beta}({\bf{q}},\omega), 
\end{equation*}

\noindent where $g_{L}$ is the Land{\'e} g-factor, the V$^{3+}$ form factor is $f(\mathbf{q})$ and $S^{\alpha\beta}(\mathbf{q},\omega)$ is the partial dynamical structure factor (with cartesian indices $\alpha,\beta$, Table \ref{Table:indices}), defined as 

\begin{equation*}
	S^{\alpha\beta}({\bf{q}},\omega)=\frac{1}{2\pi} \int dt e^{i\omega t} \langle \hat{S}^{\alpha} ({\bf{q}},t) \hat{S}^{\beta}(-{\bf{q}},0) \rangle.
\end{equation*}

\noindent The factor preceeding the partial dynamical structure factor is the polarization factor which picks out the component of the structure factor perpendicular to the scattering wavevector, to which the scattered neutrons are sensitive. We note that neutron scattering is sensitive to magnetic correlations through interacting with the spin.  It is through the presence of spin-orbit coupling in the magnetic Hamiltonian described above, neutrons are sensitive to orbital effects.  The structure factor can be related to the Green's function by way of the fluctuation-dissipation theorem

\begin{equation*}
	S^{\alpha \beta}({\bf{q}},\omega)=-\frac{1}{\pi} \frac{1}{1-\exp(-\omega/k{\rm{_{B}}}T)} \Im{G^{\alpha \beta} (\bf{q},\omega)}.
\end{equation*}

\noindent The Green's function for multi-level spin-systems can be written as a Dyson equation where the propagator describes the dynamics of the single ion at mean-field level with the inter-ion interaction treated at one-loop level with the self energy $\underline{\underline{\mathcal{J}}}(\mathbf{q})$. The single ion Green's function can be written as 

\begin{equation}
	g_{\tilde{\gamma}\tilde{\gamma}'}^{\alpha\beta}(\omega)=\sum_{qp}\frac{S^{\tilde{\gamma}}_{\alpha qp}S^{\tilde{\gamma}'}_{\beta pq}\phi_{qp}}{\omega-(\omega_{p}-\omega_{q})},
	\label{SingleionGreensFunction}
\end{equation}

\begin{figure*}
	\begin{center}
		\includegraphics[trim={0mm 20mm 0mm 0mm},width=\linewidth]{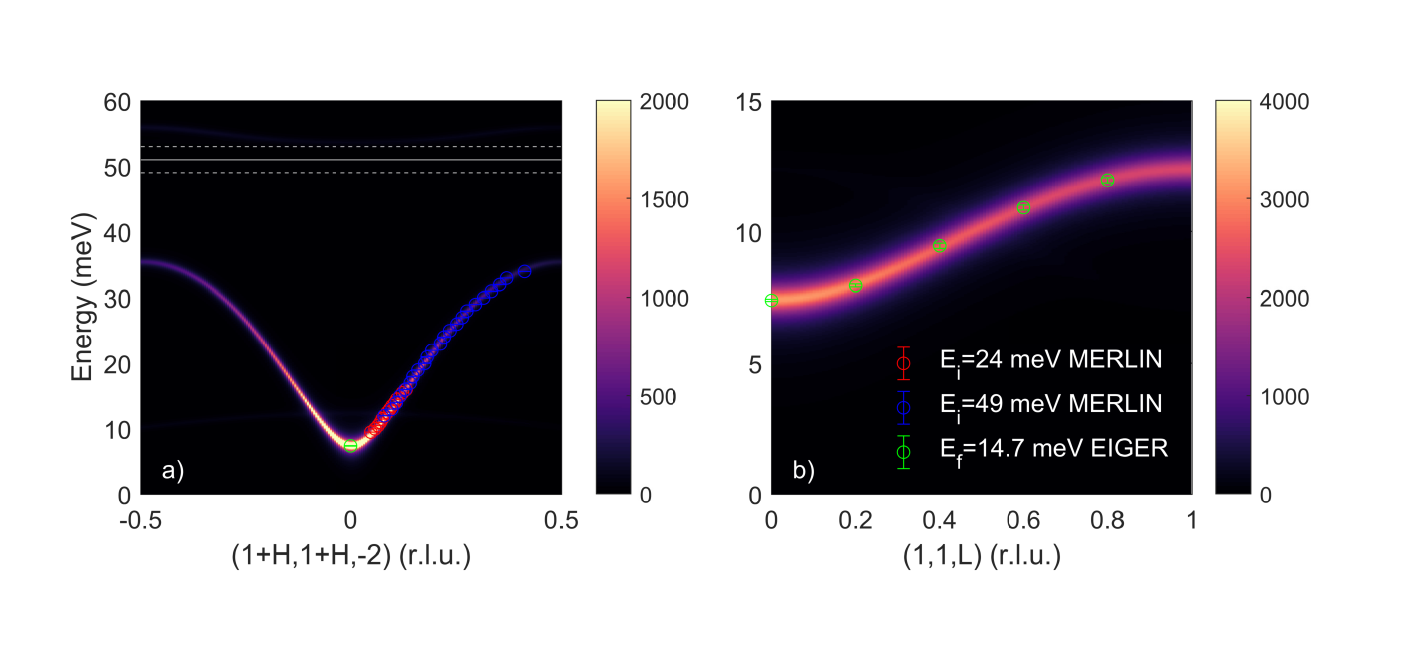}
	\end{center}
	\caption{Calculation of the component of the structure factor $a)$ parallel and $b)$ perpendicular to the chain direction. Overplotted is the extracted dispersion from neutron scattering. The green data points are from the EIGER data set and the red and blue points are from the MERLIN $E_{i} = $ 24 meV and 49 meV data sets respectively. The white solid line indicates the approximate peak position of the weak in intensity high energy mode and the dashed lines indicate the approximate width.}
	\label{fig:Theory_figure_3}
\end{figure*}

\noindent where $\omega_{p}$ is the single-ion eigenvalue of state $\ket{p}$ and $S^{\tilde{\gamma}}_{\alpha qp}=\bra{p}\hat{S}_{\alpha}^{\tilde{\gamma}}\ket{q}$. $\phi_{qp}=(f_{q}-f_{p})$ where $f_{q}$ is the Bose occupation factor of level $q$. The site indices are denoted as $\tilde{\gamma},\tilde{\gamma}'$ (Table \ref{Table:indices}). We perform a coordinate transformation onto the rotating frame so that the molecular mean field is non-oscillatory. In the rotating frame, the single-ion of each of the magnetic ions is independent of the moment direction and varies only between crystallographically inequivalent sites. In the rotating frame the Green's function is defined as

\begin{equation}
	\begin{split}
		&\tilde{G}_{\tilde{\gamma}\tilde{\gamma}'}^{\alpha\beta}(\mathbf{q},\omega)=g_{\tilde{\gamma}\tilde{\gamma}'}^{\alpha\beta}(\omega)\delta_{\tilde{\gamma}\tilde{\gamma}'}\\ &\qquad+\sum_{\gamma'}^{\mu\nu}\tilde{\mathcal{J}}^{\mu\nu}_{\tilde{\gamma}\gamma'}(\mathbf{q})g_{\tilde{\gamma}\tilde{\gamma}}^{\alpha \mu}(\omega)\tilde{G}_{\gamma'\tilde{\gamma}'}^{\nu\beta}(\mathbf{q},\omega)
	\end{split}
	\label{fulleqRF}
\end{equation} 

\noindent which can be solved as a matrix equation. Eq. \ref{fulleqRF} contains the Fourier transformed exchange coupling in the rotating frame, $\tilde{\mathcal{J}}^{\mu\nu}_{\tilde{\gamma}\gamma'}(\mathbf{q})$, which takes the form of a matrix of dimension $3N\times3N$, where $N$ is the number of sites in the unit cell. At this point it is useful to examine the structure of MgV$_{2}$O$_{4}$. Below T$_{N} \simeq$ 40 K, long-range antiferromagnetic magnetic order is established, with $\mathbf{Q}=(0,0,1)$~\cite{Wheeler10:82}.  The full crystallographic and magnetic structure can be described by an eight site unit cell as considered in Ref~\onlinecite{Perkins07:76}.

\begin{table}[h]
	\caption{\label{Unitcell} V$^{3+}$ ion positions in the unit cell.}
	\begin{ruledtabular}
		\begin{tabular}{cccc}
			Index & Moment orientation & Position vector \\ 
			\hline
			1 &$\uparrow$ & (0.125,0.625,0.125)  \\ 
			2 &$\downarrow$ & (0.375,0.875,0.125) \\
			3 &$\uparrow$ & (0.375,0.625,0.375) \\ 
			4 &$\downarrow$ & (0.125,0.875,0.375) \\
			5 &$\uparrow$ & (0.375,0.375,0.625)  \\ 
			6 &$\downarrow$ & (0.625,0.625,0.625) \\
			7 &$\downarrow$ & (0.375,0.125,0.875) \\
			8 &$\uparrow$ & (0.125,0.375,0.875) \\ 
		\end{tabular}
	\end{ruledtabular}
\end{table}   

In terms of this unit cell (Table \ref{Unitcell}), no rotation between neighboring unit cells is required and, assuming Heisenberg coupling, in the rotating frame one has 

\begin{equation}
	\underline{\underline{\tilde{\mathcal{J}}}}^{\mu\nu}_{\gamma\gamma'}(\mathbf{q})= X'\left (\underline{\underline{\tilde{\mathcal{J}}}}_{\gamma\gamma'}(\mathbf{q}) \otimes \mathbb{I}_{3}\right)X
\end{equation}

\noindent where $X=\mathrm{diag}(R_{1},...,R_{n})$ rotates the spins within the unit cell. The matrix $R_{n}$ is the $3\times 3$ rotation matrix which rotates the spin on site $n$ onto the $\hat{z}$-axis. Similarly $X'=\mathrm{diag}(R^{T}_{1},...,R^{T}_{n})$. Finally, the Green's function in the laboratory frame is found by rotating the Green function 
back using the matrices, $X, X'$

\begin{equation*}
	G_{\tilde{\gamma}\tilde{\gamma}'}^{\alpha\beta}(\mathbf{q},\omega)=X \tilde{G}_{\tilde{\gamma}\tilde{\gamma}'}^{\alpha\beta}(\mathbf{q},\omega)X'.
\end{equation*}

\noindent We now apply this theory to the neutron scattering data presented in Sect. \ref{Sect:Excitations}.

\subsection{Application to MgV$_{2}$O$_{4}$}

The excitonic dispersion relation was found according to the model presented in the previous section by finding the poles of the Green's function, $G_{\tilde{\gamma}\tilde{\gamma}'}^{\alpha\beta}(\mathbf{q},\omega)$. In Fig. \ref{fig:Theory_figure_1} $(d)$, we illustrate dipole allowed transitions through the red lines when investigating the effect of $\mathcal{H}_{MF}$.  Based on this, we would expect two branches with a lower strongly dispersive branch corresponding to transitions within the groundstate $j_{eff}$=2 manifold and another transition from the ground state to a $j_{eff}$=1 level.  We might expect that given that this is a ``crystal-field-like'' transition, this second higher energy and dipole allowed transition will be weakly dispersive in momentum akin to crystal field transitions commonly observed for rare earth ions.  

\begin{figure}
	\begin{center}
		\includegraphics[width=\linewidth]{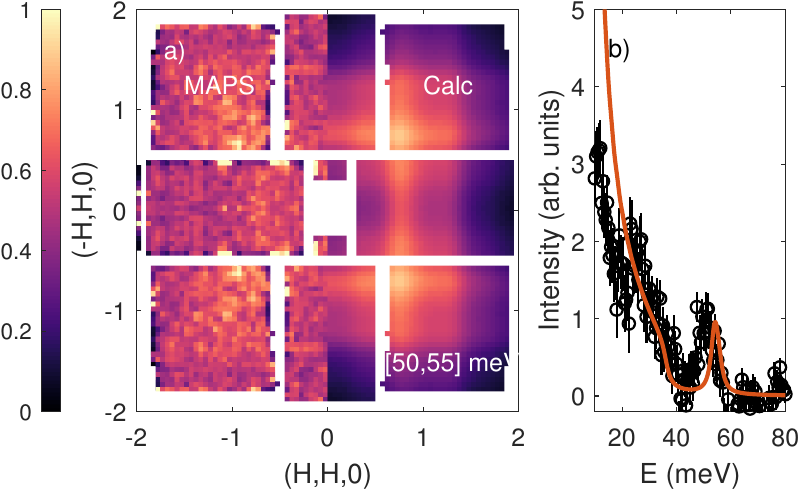}
	\end{center}
	\caption{$a)$ Constant energy slice through the inter-multiplet mode at $E\sim 50$ meV. $b)$ One-dimensional cut through $Q=(2,0)$, with multi-level spin wave calculation plotted in red. The Lorentzian half-width has been chosen to approximately match the energy resolution on MAPS, $\epsilon = 1.5$. The calculation has been integrated along the $c^{*}$ direction across the full width of the Brillouin zone. The presence of domain coexistence and the coupling of energy transfer and the wavevector along $c^{*}$ have been neglected.  The differences between calculations and data at low energies originates from an over subtraction of the data with the chosen background.}
	\label{fig:Theory_figure_4}
\end{figure}

\begin{figure}
	\begin{center}
		\includegraphics[width=\linewidth]{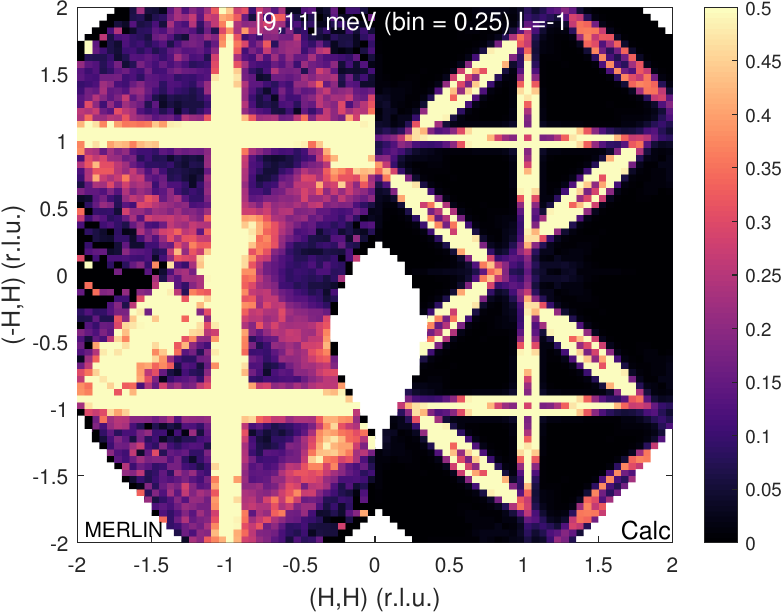}
	\end{center}
	\caption{Constant energy slice at $10\pm1$ meV, showing the comparison between the measured data from MERLIN and the excitonic model with all three domains summed as described in the text. The intensities have been integrated along the $c^{*}$ direction ($L=-1\pm 0.2$ r.l.u).}
	\label{fig:Theory_figure_5}
\end{figure}

The experimentally observed dispersion was found both along the V-V chain direction and perpendicular to the chain by fitting Gaussian peaks to one-dimensional constant energy cuts through the neutron scattering data.  These data points were then fitted using the dispersion derived according to the excitonic model in order to extract physical parameters.  The data in comparison to theory is summarized through constant momentum and energy slices in Figs. \ref{fig:Theory_figure_3}-\ref{fig:Theory_figure_5}.  The excitonic theory qualitatively shows two sets of modes with low-energy fluctuations corresponding to transitions within the ground state $j_{eff}=2$ manifold and another very weak, and comparatively disperisonless, high energy mode corresponding to transitions to the excited $j_{eff}=1$ spin-orbit manifold.  We have plotted the weak mode on the same linear intensity scale in Fig. \ref{fig:Theory_figure_3} to illustrate the strong difference in intensity between these two branches of excitations. We note that the excitonic theory predicts a very weak and comparatively flat mode at $\sim$ 10 meV.  Due to the presence of overlapping domains and a comparable magnetic zone center excitation gap, we were not able to reliably resolve this mode from our data.  

The inter-chain bond, $J_{1}$, was determined to be negligible since the inclusion of this bond (in the presence of finite $J_{3}$) gives rise to a splitting of the lower dispersive mode at approximately $Q\geq\pm0.1$ (r.l.u) which is not seen in the data. We therefore determine the magnitude of $J_{1}$ to be negligible given the absence of any of these signatures, but note that frustrated nature of this bond makes its determination difficult experimentally. The neglect of $J_{1}$ is further justified given that previous calculations have suggested that this bond is weakly ferromagnetic or weakly antiferromagnetic in the limits of ROO and COO respectively~\cite{DiMatteo05:72}. In fact, the splitting caused by $J_{1}$ is a hybridization with the lifted zero energy mode (which acquires some dispersion due to finite $J_{3}$ ) and the low energy mode due to the intra-chain coupling. As pointed out by \cite{Perkins07:76}, the presence of a zero energy mode corresponds to the degeneracy associated with the rotation of the chains in the $xy$-plane. This degeneracy is lifted by the anisotropy provided by spin orbit coupling and crystallographic distortions. A small, but finite, $J_{3}$ is expected based on the dispersive excitation seen in Fig. \ref{fig:dispersion_lowE} $(b)$ which is consistent with coupling of parallel chains. We note that there exists a further neighbor bond $J_{4}$ with bond length marginally longer than $J_{3}$. Inclusion of this bond again splits the lower dispersive mode but does not appreciably change the bandwidth or gap, provided that it is small. In the absence of any observed splitting, and in the spirit of writing down a minimal model, we neglect this term.   

\begin{table}[h]
	\caption{\label{Bond:Table} Spin-orbit exciton parameters}
	\begin{ruledtabular}
		\begin{tabular}{cccc}
			Parameter & Bond Distance (\AA) & Value (meV) \\ 
			\hline
			$J_{1}$ & 2.885 & $\approx 0$ \\ 
			$J_{2}$ & 2.885 & 19.04 ($\pm 2.86$)  \\ 
			$J_{3}$ & 4.997 & -0.173 ($\pm 0.026$) \\ 
            \hline
			$\alpha\lambda$ & - & -8.8 ($\pm 1.3$) \\ 
			$\Gamma$ & - & -42.2 ($\pm 6.33$) \\ 
			$\Theta$ & - & 0.8334 ($\pm 0.1250$) \\ 
		\end{tabular}
	\end{ruledtabular}
\end{table}   

For the lower energy dispersive mode, the data were fitted by taking one-dimensional constant energy cuts through the neutron data to extract a dispersion curve to which the poles of the Green's function were fit as the parameters were varied (Fig. \ref{fig:Theory_figure_3}). To fit the higher energy mode at $\sim 50$ meV (Fig. \ref{fig:Theory_figure_4}), a constant $\mathbf{q}$ cut was made at (2,0,0) and subtracted as a background.  This method is justified at high energy transfers hear the $\sim$ 50 meV mode, however it cuts through magnetic intensity at lower energies and therefore is an overestimate of the background.  This is reflected in Fig. \ref{fig:Theory_figure_3} which shows an overestimate of the intensity by the exciton model. The fitted parameters are listed in Table \ref{Bond:Table}. Stated uncertainties represent a nominal 15\% error to reflect the inherent uncertainties associated with the experimental measurements (such as spectrometer resolution) rather than the error bars associated with the numerical fitting procedure which are unphysically small. The approximate magnitude of the intra-chain coupling $J_{2}$ is consistent with that theoretically predicted in Ref. \onlinecite{Perkins07:76} using a Kugel-Khomskii model~\cite{Kugel82:25}. A reduction of the molecular mean field strength from that expected for an $S=1$ antiferromagnet

\begin{equation}
	\mathcal{H}_{MF}=h_{MF}\hat{S}_{z}=\Theta\left(-2J_{2}+4J_{3}\right),
\end{equation}

\noindent where $\Theta<1$, is present due to quantum fluctuations. A reduction factor of $\Theta \approx 0.9$ is expected based on the observed magnetic moment~\cite{Wheeler10:82} and the orbital projection factor based on the free ion values~\cite{Abragam:book}.  This parameter is fixed through a comparison between the calculated magnetic moment (discussed above) and the measured magnetic moment.  The calculated magnetic moment is sensitive to the projection factor $\alpha$ which is tied to the single-ion parameters $Dq$ and Racah parameter $B$.   There is, however, a degree of uncertainty about the single-ion parameters, $B$, $C$ and $Dq$ in the crystal environment, hence $\Theta$ was refined in our analysis of the neutron spectroscopy results. The refined value of $\Theta$ implies that the true value of the orbital projection factor is reduced from that expected in a free ion.  One possible origin of this is the presence of a nephelauxetic effect which reduces the electron Coulomb repulsion and hence reduces the value of the Racah $B$ parameter. 
  We note that reductions of $\sim$ 10-20 \% have been reported~\cite{Housecroft:book}, consistent with the reduction that would be required to explain the value of $\Theta$ fitted here. 

The calculated structure factor is plotted in Fig. \ref{fig:Theory_figure_3} $(a-c)$. Overplotted are the extracted peak positions from the neutron scattering data taken on EIGER and MERLIN. The calculation well describes both the dispersion and the intensity variation throughout the Brillouin zone.

In the experimental section above it was noted that single crystal samples should be expected to exhibit domain coexistence as was observed in Ref. \onlinecite{Niitaka13:111}. This occurs below the structural transition and results from the overall crystal structure retaining the average high temperature cubic symmetry in the absence of strain.  We now briefly note the consequences of domain coexistence on the neutron scattering spectrum. A rotation matrix can be defined for each domain which transforms between the crystallographic basis vectors of each domain, $d$,

\begin{equation}
	\underline{\underline{R}}_{d}\mathbf{\hat{Q}}_{0}=\mathbf{\hat{Q}}_{d}
\end{equation}

\noindent with the $\mathbf{\hat{Q}}_{0}$ defining the nominal reference basis plotted in Fig. \ref{fig:Theory_figure_1}. The crystallographic structure of MgV$_{2}$O$_{4}$ comprises chains along $[110]$ and so one can define the rotation matrices

\begin{subequations}
	\begin{gather}
		\underline{\underline{R}}_{0}=\begin{pmatrix}
			1& 0 & 0\\
			0& 1 & 0\\
			0& 0 & 1\end{pmatrix}\\
		\underline{\underline{R}}_{1}=\begin{pmatrix}
			1& 0 & 0\\
			0& 0 & 1\\
			0& 1 & 0\end{pmatrix}\\
		\underline{\underline{R}}_{2}=\begin{pmatrix}
			0& 0 & 1\\
			0& 1 & 0\\
			1& 0 & 0\end{pmatrix}
	\end{gather}
\end{subequations}

\noindent corresponding to three inequivalent domains. Fig. \ref{fig:Theory_figure_5} shows a constant energy slice with the scattering for all three domains superposed. The diagonal modes originate from the domains with basis vectors $\mathbf{\hat{Q}}_{1}$ and $\mathbf{\hat{Q}}_{2}$ with the horizontal and vertical lines originating from the $\mathbf{\hat{Q}}_{0}$ domain. A small difference in the position of the diagonal modes likely originates from the fact that the unit cell is tetragonal with $c/a = 0.994$~\cite{Wheeler10:82}. In the calculation, the domain population has been assumed to be equal however the relative intensities of the streaks in the data indicate that this is not the case. This is in agreement with previous diffraction data~\cite{Niitaka13:111}. 

\subsection{Orbital order}
\label{Sect:Orbital order}

There has been a great deal of intrigue in the nature of the orbital order in MgV$_{2}$O$_{4}$. Much of this has centered around whether the system exhibits real or complex orbital order. We will briefly summarize both orbital ordering schemes, remarking on how our results influence this discussion. 

We begin by considering the isolated single ion in an octahedral environment, neglecting the inter-ion coupling. An orbitally degenerate crystal field ground state can lower its energy via an octahedral distortion~\cite{Jahn37:161} (the Jahn-Teller effect). An elongation of the octahedron lowers the $d_{xz}$ and $d_{yz}$ orbitals by $\frac{1}{2}E_{JT}$ and raises $d_{xy}$ by $E_{JT}$~\cite{Khomskii:book}. The lower energy doublet is then split by spin-orbital coupling into $\ket{l_{z}=\pm1}=\frac{1}{\sqrt{2}}(d_{xz}\pm id_{yz})$ by $\pm \frac{1}{2}\alpha\lambda$. For a $d^{2}$ ion, assuming the spin-orbit coupling does not lift $\ket{l_{z}=-1}$ above the $d_{xy}$ orbital, then both $\ket{l_{z}=-1}$ and $\ket{l_{z}=1}$ are occupied and the resultant ground state has quenched orbital angular momentum.


Conversely, if the octahedron is compressed, the Jahn-Teller spectrum is inverted and the $d_{xy}$ level is lowered, with the $d_{xz}$ and $d_{yz}$ levels lying at $+\frac{1}{2}E_{JT}$. The ground state thus has one electron in the $d_{xy}$ orbital and one in the $\ket{l_{z}=+1}$ level. In this case the orbital angular momentum is unquenched owing to the orbital degeneracy which is broken by spin-orbit coupling, lowering the ground state energy. Based on single-ion physics alone, an elongation is favored if $|\alpha\lambda|>E_{JT}=\frac{2}{3}\Gamma$. From the fitted values in Table \ref{Bond:Table} it is clear that the compressed structure is energetically favored by the comparatively large value of $\Gamma$, hence the parameters extracted from the neutron scattering measurements are consistent with the refined crystallographic structure.

We now turn to the effect of orbital order on the inter-ion exchange in MgV$_{2}$O$_{4}$. Di Matteo \textit{et al.} and Perkins \textit{et al.} discussed this in Refs. \onlinecite{DiMatteo05:72,Perkins07:76} by way of a Kugel-Khomskii model~\cite{Kugel82:25} to model the superexchange. Here we only quote the key results. Two states were found to be consistent with the observed magnetic and crystallographic structure. The energy of the COO state is $E_{COO}=-\tilde{J}_{1}-\frac{1}{2}\tilde{J}_{2}(3+2S^{2})-\lambda-\frac{1}{2}E_{JT}$, where $\tilde{J}_{1}=J(1-\eta)/(1-3\eta)$, $\tilde{J}_{2}=J(1+\eta)(1+2\eta)$, $J=t^{2}/U_{1}$ and $\eta=J_{H}/U_{1}$. In terms of these variables, $J_{2}$=$\tilde{J}_{2}$ and, assuming we have COO, $J_{1}=\tilde{J}_{2}-2\tilde{J}_{0}$, with $\tilde{J_{0}}=\eta J/(1-3\eta)$. We thus find that our extracted exchange parameters suggest $\eta\approx 0.19$ and $J\approx 22.1$ meV which are close to values suggested by photoemission spectroscopy ($\eta_{exp}=0.11$, $J_{exp}=20.4$ meV)~\cite{Perkins07:76,Takubo06:74}. In contrast, for ROO one has $J_{1}=-\tilde{J}_{0}$, such that for $J_{1}\approx 0$ one expects $\eta \approx 0$. Such a situation is clearly unphysical as it suggests either a vanishing Hund's coupling or extremely large Coulomb term. For reasonable values of $\eta \approx 0.1$ and with $J\approx 20$ meV, the ROO model predicts $|J_{2}| \approx$ 3 meV, which would give rise to a splitting of the low energy dispersive mode that would likely be resolvable in neutron scattering measurements. 
  
  \begin{figure*}
  	\begin{center}
  		\includegraphics[width=190mm,trim=4cm 0cm 0cm 0cm,clip=true]{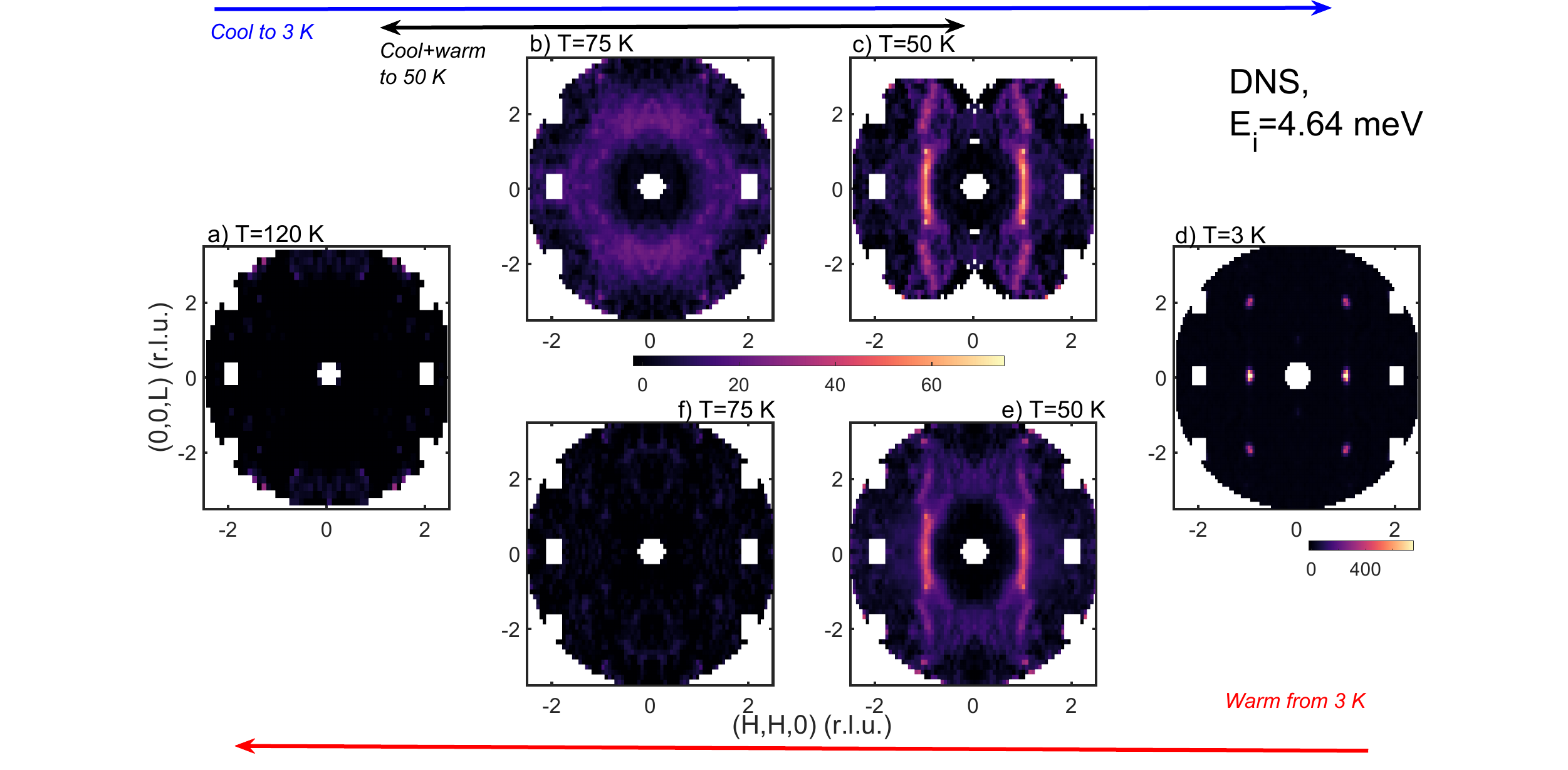}
  	\end{center}
  	\caption{The energy integrated magnetic cross section extracted from the scattering of polarized neutrons using the DNS diffractometer (FRM2).  The arrows at the top and bottom of the figure indicate the cooling and heating history. Upon cooling to 3 K and reheating the diffuse scattering evolves hysteretically with no return to the short range ordered phase at T =75 K. No such hysteresis is found if the sample is only cooled to 50 K and reheated (black arrow).  }
  	\label{fig:DNS_diffuse}
  \end{figure*}
  
As shown by Perkins \textit{et al} in Ref. \onlinecite{Perkins07:76}, qualitative differences exist between the spectra expected for ROO and COO, namely an optical mode is present in the case of COO whereas ROO gives rise only to an acoustic mode. This optical mode finds its origin in the unquenched orbital angular momentum which permits inter-multiplet transitions between the $j_{eff}=2$ and $j_{eff}=1$ levels (Fig. \ref{fig:Theory_figure_1} $d$). Our neutron scattering experiments confirm the existence of this optical mode at $E\sim 50$ meV which we were able to model within a multi-level excitonic formalism. We find that the resultant single ion ground state, consistent based on the fitted parameters (Table \ref{Bond:Table}), is the $\ket{l_{z}=1,S_{z}=1}$ state presented in Ref. \onlinecite{DiMatteo05:72}. In this state, the effective orbital angular momentum aligns with the spin moment on all V$^{3+}$ sites. It should be noted that due to the negative projection factor, this corresponds to the antialignment of the spin and \textit{true} orbital moments which gives rise to the reduction in the magnetic moment measured in diffraction~\cite{Wheeler10:82}. By treating the spin-orbit coupling, tetragonal distortion and molecular field on the same level using an excitonic model, we have demonstrated that MgV$_{2}$O$_{4}$ can be understood to behave according to the the COO picture presented in Refs. \onlinecite{DiMatteo05:72,Perkins07:76} though the distortion, spin-orbit coupling and molecular mean field strengths are comparable. This analysis helps to explain the apparent discrepancy between the previous neutron scattering results~\cite{Wheeler10:82} and the predicted spectrum for COO~\cite{Perkins07:76} which considered $|\Gamma|\ll|\lambda|$.

We note that Di Matteo \textit{et al.}~\cite{DiMatteo05:72} propose several types of COO ranging from all V$^{3+}$ being in a complex orbital state and a mixture between real and complex orbital orders  on different sites. In our theoretical analysis discussed above, we have considered the case that all V$^{3+}$ ions are in a complex orbital state.  Qualitatively, a mixture of real and complex orbital states would decrease and likely damp the high energy spin-orbit exciton corresponding to excitations from the ground state $j_{eff}=2$ manifold to a higher energy $j_{eff}=1$ spin-orbit manifold. Given how weak this mode is this would likely not be observable. 
 Another point against a mixture of complex and real orbitals is crystal lattice symmetry with the ground state being inconsistent with such a situation.  We therefore consider here the case of full complex orbital order (COO) in MgV$_{2}$O$_{4}$.

\section{Hysteretic Magnetic correlations}
\label{Sect:Hysteresis}

There are three different spin-orbital phases in MgV$_{2}$O$_{4}$ on cooling from high temperature. $(1)$ At high temperatures exceeding T$_{c}$, the nuclear unit cell is cubic and the magnetism is paramagnetic. $(2)$ In the temperature regime T$_{N}$ $<$ T $<$ T$_{S}$, the structural unit cell distorts to tetragonal.  $(3)$ For low temperature T  $<$ T$_{N}$, antiferromagnetic order sets in.   As displayed in Fig. \ref{fig:Theory_figure_1} $(d)$, the single-ion spin-orbital ground state is different in each one of these phases transitioning from a degenerate $j_{eff}=2$ state in the high temperature cubic phase to an orbital doublet below T$_{S}$, and then finally this degeneracy being split by a molecular field in the low temperature antiferromagnetic phase. This temperature dependence opens up the possibility of hysteretic effects on cooling and warming through the intermediate phase given the orbital degeneracy present, which is broken when cooling through the antiferromagnetic transition by a Zeeman like molecular field.  We now investigate these spin-orbital ground states using energy-integrated magnetic diffuse scattering. 

To investigate the low-energy critical magnetic fluctuations as a function of temperature in MgV$_{2}$O$_{4}$, we studied the magnetic cross section using the DNS polarized diffractometer.  The combined magnetic intensities from the diffuse scattering measurements are displayed in Fig. \ref{fig:DNS_diffuse}. We note the these measurements display a hysteresis depending on if the sample is cooled below the magnetic ordering temperature, T$_{N}$ $\simeq$ 40 K. Cooling to T= 75 K from 120 K (Fig. \ref{fig:DNS_diffuse} $a\rightarrow b$) one sees the development of magnetic diffuse scattering, consistent with a spin-frustrated pyrochlore system above the structural transition.  On further cooling to T=50 K (Fig. \ref{fig:DNS_diffuse} $b\rightarrow c$), the magnetic scattering changes into chain-like rods below the structural transition. Upon heating from T = 50 K back to 75 K (Fig. \ref{fig:DNS_diffuse} $c \rightarrow b$) one recovers these short-ranged magnetic correlations. However, if the sample is cooled below the long-range magnetic transition, T$_{N}$ and then warmed up to T=75 K (Fig. \ref{fig:DNS_diffuse} $a\rightarrow d$) a different behavior is observed.  Whilst at T = 50 K (Fig. \ref{fig:DNS_diffuse} $d \rightarrow e$) one still observes chain-like scattering, above the structural transition (Fig. \ref{fig:DNS_diffuse} $e\rightarrow f$), T$_{S}$, the short-ranged correlations are absent and no magnetic scattering is observable.  This indicates that either the magnetic fluctuations have moved outside our energy window determined by kinematics on DNS with fixed E$_{i}$=4.64 meV, or the critical fluctuations are highly extended in momentum and energy.

We suggest that the explanation for this hysteresis tied to cooling through the magnetic ordering transition at T$_{N}$ lies in the orbital order present in MgV$_{2}$O$_{4}$.  At T=75 K, above the structural transition, in terms of a real orbital basis the ground state has equal occupation of $\ket{xy}$, $\ket{xz}$ and $\ket{yz}$ orbitals (Fig. \ref{fig:Theory_figure_1}).  In this state the strength of the superexchange is the same for both inter and intra-chain bonds, $J_{1}=J_{2}$ and MgV$_{2}$O$_{4}$ is a canonical frustrated pyrochlore. As one cools below the structural transition, the local octahedra tetragonally compress leading to a doubly degenerate state with $l_{z}=\pm 1$. The average occupancy of both the $\ket{xz}$ and $\ket{yz}$ orbitals is $\frac{1}{2}$ and so the strength of the inter-chain coupling $J_{1}$ is reduced, rendering the system quasi-one-dimensional which gives rise to rods in the magnetic diffuse scattering. At this point, long-range magnetic order has yet to be established and so the orbital ground state is doubly degenerate. Below the antiferromagnetic ordering temperature, three-dimensional order is established thanks to the cooperative effect of the third-nearest-neighbor coupling $J_{3}$ and the anisotropy provided by the distortion and spin-orbit coupling.


One key effect of longer-range coupling (beyond $J_{1}$ and $J_{2}$) is to reduce the degree to which the spin moment is suppressed due to fluctuations, both thermal and due to the logarithmic divergence of spin fluctuations in the one-dimensional quantum antiferromagnet (as per the Mermin-Wagner theorem~\cite{Mermin66:17}). Upon establishment of long-range order, the orbital doublet ground state is split by the molecular mean field, establishing the COO ground state of Refs. \onlinecite{DiMatteo05:72} and \onlinecite{Perkins07:76}. As the sample is then reheated, at T~$ >$ T$_{N}$ long-range magnetic order is lost as fluctuations build, but we speculate that the orbital moments remain frozen in their COO configuration as the tetragonal symmetry prevents them from reorienting. Finally as one reaches the structural transition from below, it should be expected that the orbital moments reorient. However,  given the frozen nature of the orbital moments in the ground state established by magnetic order, this is prevented by a potential barrier established by the low temperature molecular fields.  To overcome this requires a thermal energy equivalent to the molecular field of order $\sim$ 10 meV, or above $\sim$ 100 K.  We speculate that through the Zeeman energy applied by the molecular field induced through magnetic  ordering, MgV$_{2}$O$_{4}$ displays an orbital memory of the low temperature spin order.

\section{Conclusions}

In conclusion, we have mapped out the spin-orbital fluctuations in MgV$_{2}$O$_{4}$ using neutron spectroscopy where we have observed two different branches.  We have parameterized these in terms of a spin-orbital excitonic theory where the single-ion Hamiltonian is used to determine the quantized ground state which is then coupled on a lattice using RPA.  The results are strongly supportive of a COO ground state.  We then used this model to understand hysteretic magnetic fluctuations through the three different spin-orbital phases.


\begin{acknowledgements}

The authors thank E. Chan and M. Mourigal for useful discussions. H. L. was co-funded by the ISIS facility development studentship programme and the EPSRC. C. S. acknowledges support from the Carnegie Trust for the Universities of Scotland, EPSRC, and the STFC.  P.M.S. acknowledges support from the California NanoSystems Institute through the Elings Fellowship program. S.D.W. acknowledges financial support from the US Department of Energy (DOE), Office of Basic Energy Sciences, Division of Materials Sciences and Engineering under Grant No. DE-SC0017752.  The authors gratefully acknowledge the financial support provided by the J\"{u}lich Centre for Neutron Science (JCNS) to perform the neutron scattering measurements at the Heinz Maier-Leibnitz Zentrum (MLZ), Garching, Germany.

\end{acknowledgements}

\section{Appendix - sample characterization}

In this section we outline additional experimental information regarding crystal growth and characterization of the nuclear and magnetic properties of our MgV$_{2}$O$_{4}$ samples.

\subsection{Additional experimental information}

\textit{Materials Preparation-} Single crystals of MgV$_{2}$O$_{4}$ were grown using a four mirror image furnace (Crystal Systems Inc.).  Dry magnesium oxide, MgO (Sigma Aldrich, 99.995~\%) and vanadium (III) oxide, V$_{2}$O$_{3}$ (Sigma Aldrich, 99.99~\%) were ground in an agate mortar and pestle with an excess of 1.5~\% V$_{2}$O$_{3}$. The excess V$_{2}$O$_{3}$ was added to combat the evaporation of vanadium oxide during the growth. For single crystal growth, the ground powders were pressed in a rod and sintered in the image furnace on a lower power in a flowing argon atmosphere (10 bar pressure, 0.1~L/min).  The crystal growth conditions were 5~mm/hr growth speed, 35~rpm rotation, 10 bar argon with flow rate $\sim$ 0.1~L/min.  The partial oxygen pressure was monitored on the input and output of the gas supply (Cambridge Sensotec) and the flow rate was adjusted to keep the partial pressure below $\sim$10-15~ppm.  A radiation-heated molybdenum getter was installed in the furnace to help reduce the oxygen partial pressure and keep it stable over the growth.  Powders of MgV$_{2}$O$_{4}$ were synthesized using the same procedure (and samples) outlined in Ref. \onlinecite{Browne20:101}.

\textit{Polarized neutron diffraction-} To investigate the momentum-broadened magnetic correlations characterizing the critical properties in the paramagnetic phases,  polarized diffuse scattering measurements were performed on the cold neutron diffractometer DNS at the Heinz Maier-Leibnitz Zentrum (MLZ) with an incident neutron wavelength of 4.2~\AA~selected with a double-focusing pyrolytic graphite monochromator. While DNS has time-of-flight capabilities, data presented was collected in $\textit{non}$ time-of-flight mode, providing the integrated scattering intensity with energy transfers up to an E$_{i}$ = 4.64~meV. The flipping ratio was measured to be 20 and a weak 50~Gauss guide field was applied at the sample for the preservation of neutron polarization. A vanadium standard was used to account for variations of the detector response and the solid angle coverage. Corrections for the polarization efficiency of the supermirror analyzers were performed by using the scattering from a NiCr alloy. Each measurement was performed over a period of 8-12~hours after a sample thermalization period of 1~hour. 

\textit{Unpolarized neutron diffraction-} To characterize the structural properties of our single crystals in both the cubic and tetragonal phases, unpolarized single crystal neutron diffraction data were taken on the D9 diffractometer (ILL, France) with a fixed incident wavelength of $\lambda$=0.842 \AA. 

\textit{Synchrotron x-ray diffraction-} Temperature dependent x-ray diffraction on powders were performed to confirm structural properties and to search for any hysteretic effects in the lattice properties.  Experiments were performed on the I11 high resolution beamline~\cite{Thompson09:80} at the Diamond Light Source (Didcot, UK) using the MAC detector system.  A wavelength of $\lambda$ = 0.8258\AA~was used.

\subsection{Thermodynamics}

\begin{figure}
	\begin{center}
		\includegraphics[width=95mm,trim=4cm 6cm 2cm 6cm,clip=true]{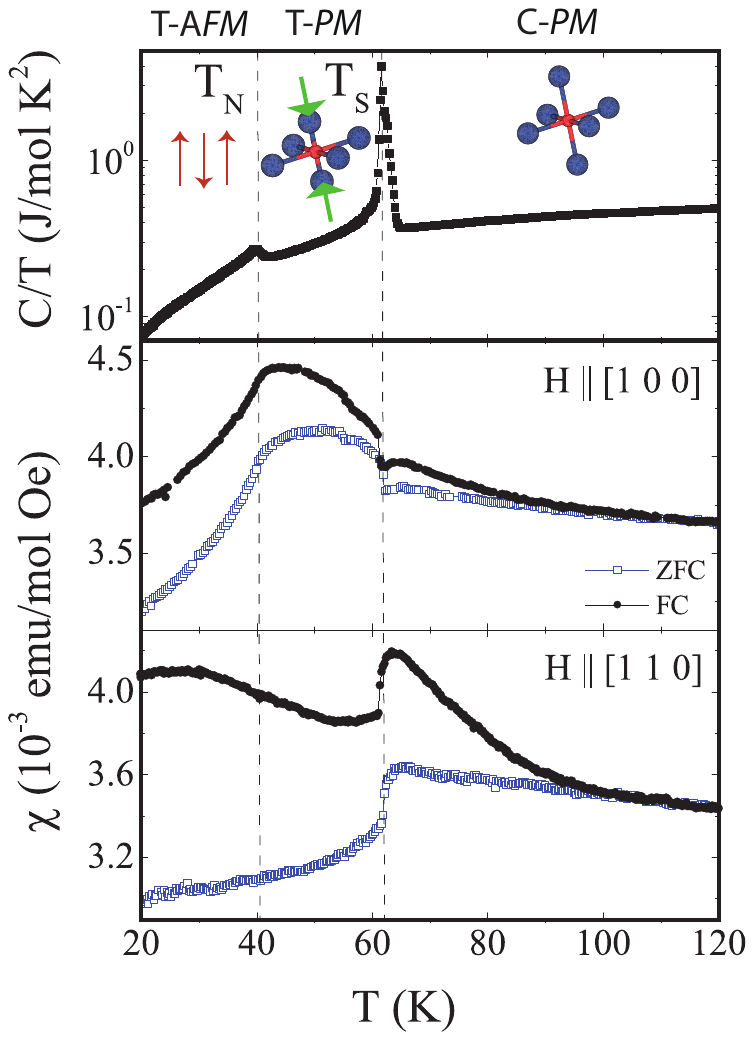}
	\end{center}
	\caption{The heat capacity and magnetic susceptibility measured on a piece of the single crystals used for our neutron studies and displayed in Fig. 1.  The data clearly show the presence of two transitions which we will show below are a cubic to tetragonal structural transition ($\simeq 60 K$) and a lower antiferromagnetic transition ($40 K$) based on neutron and x-ray diffraction.}
	\label{fig:hc_chi}
\end{figure}

Heat capacity and magnetic susceptibility measurements (taken using Quantum Design MPMS and PPMS systems) on a small piece cut from the crystals (Fig. \ref{fig:figure1} $(b)$) used for our neutron studies are displayed in Fig. \ref{fig:hc_chi}.  The data display two transitions at $\simeq$ 60 K and 40 K.  We discuss below based on neutron diffraction on pieces of these crystals that these two transitions represent a high temperature structural transition from cubic to tetragonal unit cells (T$_{S}\simeq$60 K) and a lower temperature transition to an antiferromagnetically ordered phase (T$_{N}\simeq$ 40 K).  

\subsection{Temperature dependent diffraction}

\begin{figure}
	\begin{center}
		\includegraphics[width=75mm,trim=0cm 0cm 0cm 0cm,clip=true]{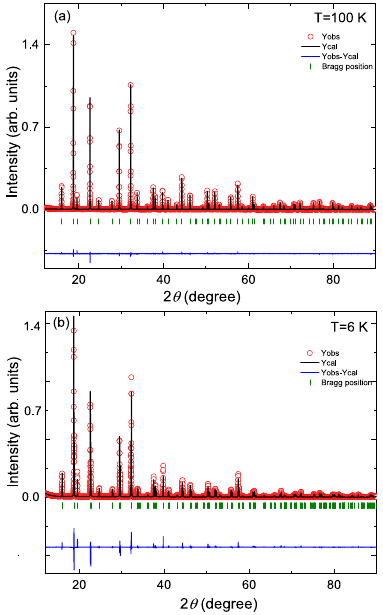}
	\end{center}
	\caption{Representative refinements of MgV$_{2}$O$_{4}$ powders used to characterize the structural transition in the $(a)$ cubic phase at T=100 K and in the $(b)$ tetragonal phase at T=6 K.}
	\label{fig:synchrotron1}
\end{figure}

\begin{figure}
	\begin{center}
		\includegraphics[width=87mm,trim=1cm 1.0cm 1cm 0cm,clip=true]{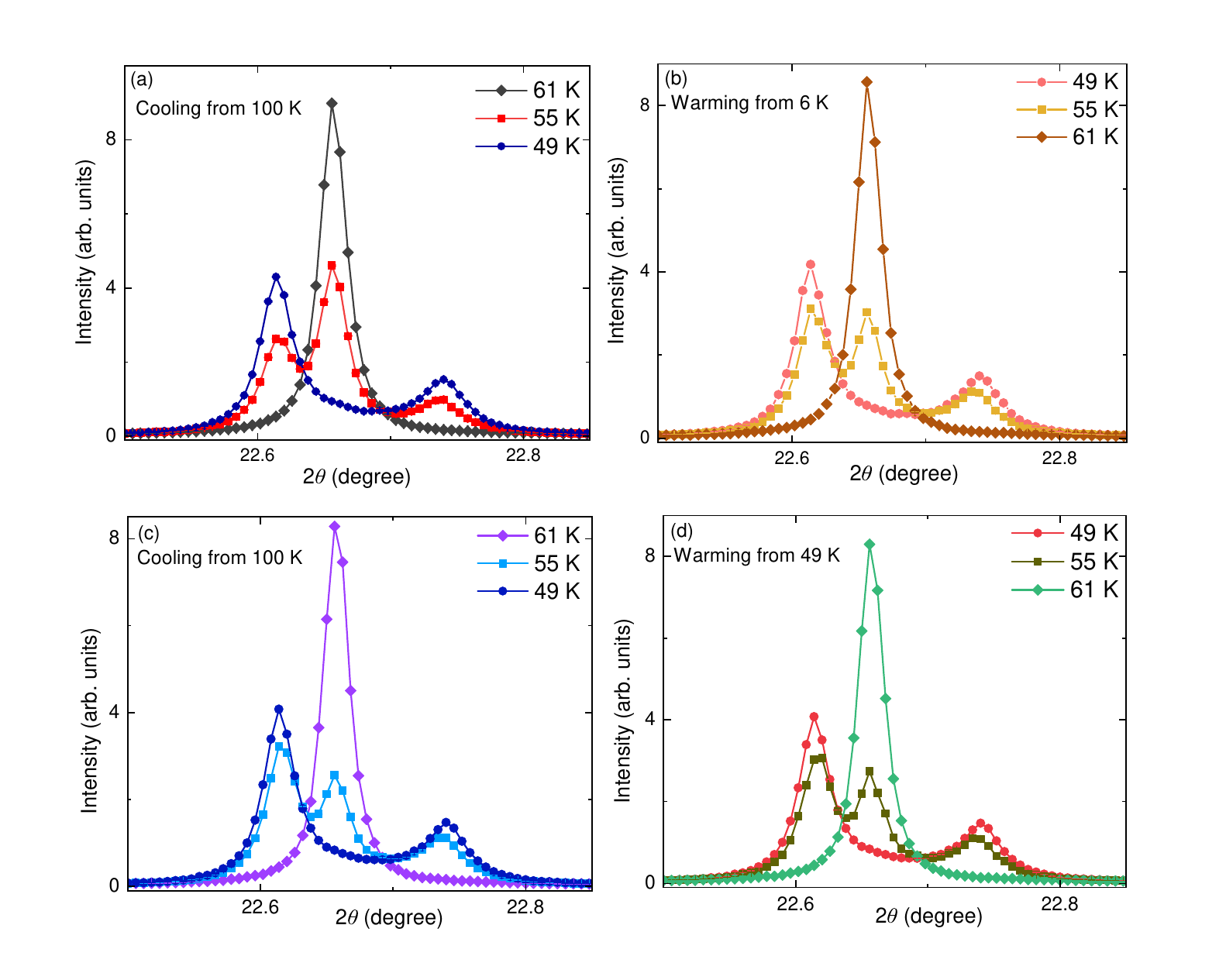}
	\end{center}
	\caption{Representative scans of the structural cubic (4,0,0) Bragg peak near the structural transition temperature.  Different temperature cycles are displayed.  The temperature cycle is $(a) \rightarrow$ $(b) \rightarrow$ $(c) \rightarrow$ $(d)$.}
	\label{fig:synchrotron2}
\end{figure}

We discuss the two phase transitions present in our single crystals using temperature dependent powder and single crystal diffraction.  In terms of single crystal neutron diffraction, Fig.~\ref{fig:figure1}$ (c)$ in the main paper illustrates the (0,0,2) intensity on cooling and warming displaying a structural change at $\simeq$ 60~K with a significant difference in the intensity on heating and cooling.  We speculate this originates from domains and strain in the sample resulting in changes in extinction on cooling/warming.   We further discuss this point with regards to synchrotron x-ray measurements below.  Fig. \ref{fig:figure1} $(c)$ also plots the onset of magnetic order at T$\rm{_{N}} \simeq$ 40~K probed with neutron diffraction by monitoring the (1,1,0) magnetic Bragg peak.  The results are consistent with published data and illustrative of the sample quality.~\cite{Watanabe14:90,Islam12:85,Mamiya97:81,Wheeler10:82,Watanabe14:90,Islam12:85}  

We further studied the structural transition using synchrotron x-ray diffraction and Fig.~\ref{fig:figure1}$ (c)$ displays the lattice parameters as a function of temperature.  This further demonstrates that the high temperature transition is structural.  We have confirmed the unit cells and structural parameters in the literature in both the high and low temperature phases and representative diffraction patterns are shown in Fig. \ref{fig:synchrotron1}.

An interesting result in the context of our hysteretic studies outlined in the main text is shown in Fig. \ref{fig:synchrotron2} where we plot representative powder data of the (4,0,0) Bragg peak (in the cubic setting).  At high temperatures, a single peak is displayed representative of the cubic unit cell and a splitting is seen at low temperatures indicative of a structural distortion to a tetragonal unit cell.  However, for a significant temperature region below the structural transition, a coexistance is observed.  The relative fractions of the two structural phases, indicated by the relative peak intensities, is different dependent on the temperature history.  This supports that the hysteresis in intensity displayed in Fig. \ref{fig:figure1} originates from relative domain populations.

\subsection{Single Crystal Nuclear Structure}

To characterize the structural properties in our single crystals we applied neutron diffraction on a small piece from our crystals used for spectroscopy.  Owing to the large uncertainty related to possible mixing between the Mg and V sites in single crystalline \mvo, it was important to measure the nuclear structure above and below T$\rm{_{S}}$=60~K.   As demonstrated in Ref. \onlinecite{Islam12:85}, the site mixing has strong effects on the magnetic and structural transitions.  

In agreement with previous studies (Ref. \onlinecite{Niitaka13:111}), the refinements of the T=140~K and 5~K temperature (Fig. \ref{fig:figure1}) data sets confirm the high-temperature $Fd\overline{3}m$ and low temperature $I4_{1}/amd$ crystal structures for MgV$_{2}$O$_{4}$, with $a$=8.42 \AA\ (cubic phase), and $a$=5.77 \AA\ and c=8.16 \AA\ in the tetragonal phase.  The Mg and O sites were refined to be fully occupied within error, demonstrating the quality of the single crystal.  The refined structural parameters (including domains which are further discussed below) are listed in Table ~\ref{table_absolute_a}.  

There has been some discussion in the literature about the possibility of different structures based on the observation of a predicted extinct $\vec{Q}$=(0,0,2) Bragg peak.~\cite{Wheeler10:82}  We have observed this Bragg peak in our D9 unpolarized experiment.  However, an azimuthal scan keeping the momentum transfer fixed found significant intensity variations indicating that this peak is likely due to multiple scattering processes. We therefore consider the low-temperature structure to be $I4_{1}/amd$ for the purposes of this paper.

The high temperature cubic $Fd\overline{3}m$ space groups has a fourfold symmetry along each crystallographic axis.  On cooling a macroscopic crystal through the Jahn-Teller transition to a $I4_{1}/amd$ tetragonal space group with a twofold and a fourfold symmetry axis.  In the absence of an external symmetry breaking field such as strain this symmetry must, on average, be preserved, resulting in structural domains at low temperatures.  There are therefore three possible low temperature domains associated with a distortion along one of the axes of the high temperature unit cell.  These domains were refined not to be of equal population as indicated by the results in Table \ref{table_absolute}. This unequal domain population is also consistent with the findings of Ref. \onlinecite{Niitaka13:111} and may be domain and environment dependent.

\begin{table}[ht]
	\caption{Atomic parameters for MgV$_{2}$O$_{4}$ at T=140 K (cubic) and 50 K (tetragonal) from single crystal (neutron diffraction.  Positions in the cubic phase are $8a$ (${1\over8}$, ${1\over8}$, ${1\over8}$), $16d$ (${1\over2}$, ${1\over2}$, ${1\over2}$), and $32e$ ($x$, $x$, $x$) for Mg, V, and O respectively, and in the $4a$ (0, ${3\over 4}$, ${1\over 8}$), $8d$ (0, 0, ${1 \over 2}$), and $16h$ (0, $y$ , $z$) for the tetragonal phase.  V was fixed to full occupancy. $*$ Denotes parameters fixed from the T=140~K refinement.  Three different domains were observed and fractions included in the T=50 K refinement yield 88(1)\%, 8(1)\%, and 4(1)\% populations.}
	\centering
	\begin{tabular} {c c c }
		\hline\hline
		& T=140 K  & T=50 K*  \\
		\hline\hline
		Space Group & $Fd\overline{3}m$ & $I4_{1}/amd$ \\
		$x$(O) & 0.26000(6)  & 0.0    \\
		$y$(O) & - & 0.47929(14)    \\
		$z$(O) & - & 0.26002(8)    \\
		Mg $B_{iso}$ (\AA$^{2}$) & 0.51(4) & 0.16(3) \\
		O $B_{iso}$ (\AA$^{2}$) & 0.45(4) & 0.24(2) \\
		Mg occupancy (\%) & 97(7) & 97* \\
		O occupancy (\%) & 94(6) & 94* \\
		\hline
		No. reflections & 741 & 827 \\
		No. indep. ref. & 74 & 188 \\
		R$_{f}$ \% & 2.55 & 4.63 \\
		R$_{w}$ \% & 4.83 & 6.29 \\
		$\chi^{2}$ & 2.15 & 3.14 \\
		\hline
		\label{table_absolute_a}
	\end{tabular}
\end{table}

\begin{table}[ht]
	\caption{Atomic parameters for MgV$_{2}$O$_{4}$ at T=100 K (cubic) and 7 K (tetragonal) from the powder sample (synchrotron).  Positions in the cubic phase are $8a$ (${1\over8}$, ${1\over8}$, ${1\over8}$), $16d$ (${1\over2}$, ${1\over2}$, ${1\over2}$), and $32e$ ($x$, $x$, $x$) for Mg, V, and O respectively, and in the $4a$ (0, ${3\over 4}$, ${1\over 8}$), $8d$ (0, 0, ${1 \over 2}$), and $16h$ (0, $y$ , $z$) for the tetragonal phase.}
	\centering
	\begin{tabular} {c c c }
		\hline\hline
		& T=100 K  & T=7 K  \\
		\hline\hline
		Space Group & $Fd\overline{3}m$ & $I4_{1}/amd$ \\
		$x$(O) & 0.25985(4)  & -   \\
		$y$(O) & - & 0.01833(2)    \\     
		$z$(O) & - & 0.26108(11)   \\
		Mg $B_{iso}$ (\AA$^{2}$) & 0.386(5) & 0.240(11) \\
		O $B_{iso}$ (\AA$^{2}$) & 0.578(6) & 0.360(11) \\
		V $B_{iso}$ (\AA$^{2}$) & 0.366(1) & 0.234(3) \\
		\hline
		R$_{p}$ \% & 5.17 & 8.96 \\
		R$_{wp}$ \% & 7.19 & 12.7 \\
		R$_{B}$ \% & 2.28 & 5.44 \\
		R$_{f}$ \% & 1.57 & 4.15 \\
		\hline
		\label{table_absolute}
	\end{tabular}
\end{table}

\subsection{Magnetic Structure}

To confirm the magnetic structure, we have used polarized neutron scattering measurements at DNS (MLZ) in the low temperature ordered phase with the sample aligned in the (HHL) scattering plane.   The measured magnetic Bragg peaks and intensity differences are consistent with the magnetic structure reported in ZnV$_{2}$O$_{4}$ (Ref. \onlinecite{Reehuis03:35}) and is illustrated in Fig. \ref{fig:figure1} $(a)$.  To guide the analysis of the magnetic structure, we have used the program SARAh~\cite{Wills00:276} to calculate the irreducible representations and basis vectors for the space group.  There are 2 common irreducible representations for the space group $I4_{1}/amd$ with propagation vector (001).   The polarized data at low temperatures shows a uniaxial structure within experimental error consistent with only one irreducible representation being involved in the magnetic transition.  The involvement of basis functions from only one irreducible representation is consistent with the T$_{N}$ $\simeq$ 40 K magnetic transition being second order (Fig. \ref{fig:figure1} $c$) as predicted from Landau theory.   The magnetic moment was estimated by comparing magnetic $\vec{Q}$=(110) and nuclear (111) Bragg peaks to be $\mu$=0.70 $\pm$ 0.15 $\mu_{B}$. This is in reasonable agreement with the \textit{refined} magnetic moment value found by neutron diffraction in Ref. \onlinecite{Wheeler10:82}, $\mu_{z}=0.47\mu_{B}$ (with an $\sim$ 8$^{\circ}$ canting from the $z$-axis). The magnetic structure in Fig. \ref{fig:figure1} $(a)$ and Fig. \ref{fig:Theory_figure_2} is based upon interpenetrating chains of magnetic V$^{3+}$ which carry a magnetic moment.  The chain-like structure originates from the tetragonal structural distortion which occurs at T$_{s}$ that breaks the frustrating magnetic interactions present in the high temperature cubic unit cell as discussed in the main text.

\subsection{Polarization analysis}

Polarized neutron spin-flip (SF) and non-spin-flip (NSF) scattering using an $xyz$ geometry was performed on the DNS diffractometer.  Magnetic neutron scattering obeys the selection rule that it is sensitive only to components of the magnetic vector perpendicular to the momentum transfer.  In Ref. \onlinecite{Shirane:book}, this projected vector is denoted as $\vec{S}_{\perp}$  and polarized neutron scattering affords a means of separating out the various components of $\vec{S}_{\perp}$.  The components of $\vec{S}_{\perp}$ which are perpendicular to the neutron polarization appear in the SF channel, while the component parallel to the neutron polarization will produce NSF scattering.

A problem with performing polarized neutron experiments on vanadium based materials is the fact that vanadium has a strong incoherent cross section which translates into an overall momentum independent background.  In polarized experiments, the incoherent cross section in the SF channel is twice that in the NSF channel, which complicates comparing the channels directly.  To circumvent this, we considered the $xyz$ channels in the SF and NSF channels separately.  In our experiment, the $x$ channel was defined as when the neutron polarization was parallel to the average momentum transfer $\vec{Q}$; $y$ was set to be perpendicular to $x$, but within the horizontal scattering plane; and $z$ was defined as vertical being perpendicular to $x$, $y$, and $z$ is normal to the horizontal scattering plane.  

Given the selection rules by neutron scattering $S_{\perp,x}$ is always absent.  The problem is then to extract $S_{\perp,y}$ and $S_{\perp,z}$ reliably.  Using these combined selection rules we note the following for the magnetic neutron intensities ($I$) in the SF channel including incoherent background ($B_{incoh}$):

\begin{eqnarray}
	I_{SF,x}=S_{\perp,y}+S_{\perp,z} +B_{incoh,SF} \nonumber \\
	I_{SF,y}=S_{\perp,z}+B_{incoh,SF} \nonumber \\
	I_{SF,z}=S_{\perp,y} +B_{incoh,SF} 
\end{eqnarray}

\noindent and in the NSF channel,

\begin{eqnarray}
	I_{NSF,x}=B_{incoh,NSF} \nonumber \\
	I_{NSF,y}=S_{\perp,y}+B_{incoh,NSF} \nonumber \\
	I_{NSF,z}=S_{\perp,z}+B_{incoh,NSF} 
\end{eqnarray}

\noindent Since the magnetic diffuse scattering does not overlap with the nuclear scattering we have not considered feedthrough from one channel to another due to the incomplete flipping ratio or misalignment of the spin with respect to the momentum transfer.  We have removed the scattering around the nuclear Bragg peaks in this analysis.  In an ideal experimental setup, $2 \times B_{incoh,NSF}=B_{incoh,SF}$, however we have treated SF and NSF scattering separately to extract the individual cross section for $S_{\perp,y}$ and $S_{\perp,z}$.  For example, from the SF channel, $S_{\perp,y}=I_{SF,x}-I_{SF,y}$ and $S_{\perp,z}=I_{SF,x}-I_{SF,z}$.  From the the NSF channel, $S_{\perp,y}=I_{NSF,y}-I_{NSF,x}$ and $S_{\perp,z}=I_{NSF,z}-I_{NSF,x}$.  Therefore, the required components for $S_{\perp,y}$ and $S_{\perp,z}$ can be extracted independently from both the SF and NSF channels and combined.  This method provides us a means of subtracting off the incoherent background from vanadium.

\end{document}